# Simultaneous thermoosmotic and thermoelectric responses in nanoconfined electrolyte solutions: Effects of nanopore structures and membrane properties[1]

Wenyao Zhang[1,2], Muhammad Farhan[1], Kai Jiao[1], Fang Qian[1], Panpan Guo[1], Qiuwang Wang[1], Charles Chun Yang[2], and Cunlu Zhao[1, *]

[1]MOE Key Laboratory of Thermo-Fluid Science and Engineering, School of Energy and Power Engineering, Xi'an Jiaotong University, Xi'an 710049, China

[2]School of Mechanical and Aerospace Engineering, Nanyang Technological University, 50 Nanyang Avenue, Singapore 639798, Singapore

*Corresponding author. Email: mclzhao@xjtu.edu.cn (C.Z.)

---

# Abstract

*Hypothesis*: Nanofluidic systems provide an emerging and efficient platform for thermoelectric conversion and fluid pumping with low-grade heat energy. As a basis of their performance enhancement, the effects of the structures and properties of the nanofluidic systems on the thermoelectric response (TER) and the thermoosmotic response (TOR) are yet to be explored.

*Methods*: The simultaneous TER and TOR of electrolyte solutions in nanofluidic membrane pores on which an axial temperature gradient is exerted are investigated numerically and semi-analytically. A semi-analytical model is developed with the consideration of finite membrane thermal conductivity and the reservoir/entrance effect.

*Findings*: The increase in the access resistance due to the nanopore-reservoir interfaces accounts for the decrease of short circuit current at the low concentration regime. The decrease in the thermal conductivity ratio can enhance the TER and TOR. The maximum power density occurring at the nanopore radius twice the Debye length ranges from several to dozens of mW $K^{-2}$ $m^{-2}$ and is an order of magnitude higher than typical thermo-supercapacitors. The surface charge polarity can heavily affect the sign and magnitude of the short-circuit current, the Seebeck coefficient, and the open-circuit thermoosmotic coefficient, but has less effect on the short-circuit thermoosmotic coefficient. Furthermore, the membrane thickness makes different impacts on TER and TOR for zero and finite membrane thermal conductivity.

**Graphical Abstract**

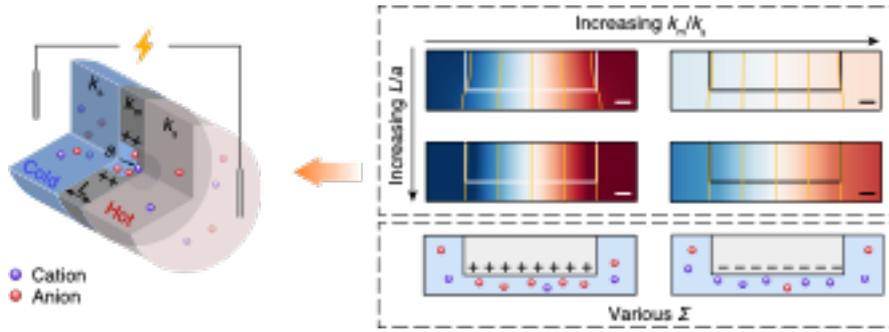

**Nomenclature**

*Abbreviations*

| | |
|---|---|
| EDL | electric double layer |
| ODE | ordinary differential equation |
| OCV | open circuit voltage |
| PDE | partial differential equation |
| SCC | short-circuit current |
| TER | thermoelectric response |
| TOR | thermoosmotic response |
| TOC | thermoosmotic coefficient |

*Latin symbols*

| | |
|---|---|
| $a$ | nanopore radius (m) |
| $a_{res}$ | reservoir radius (m) |
| $A$ | cross-section area of nanopore (m$^2$) |
| $c_p$ | heat capacity at constant pressure (J K$^{-1}$ kg$^{-1}$) |
| $D_i$ | diffusion coefficient of ionic species $i$ (m$^2$ s$^{-1}$) |
| $e$ | elementary charge (C) |
| $F$ | Faraday constant (C mol$^{-1}$) |
| $I$ | electric current (A) |
| $I_0$ | reference current defined as $I_0 = \pi a^2 e u_{ref} n_0$ (A) |
| $\boldsymbol{J_i}$ | flux of ionic species $i$ (m$^2$ s$^{-1}$) |
| $J_{ion}$ | average ionic flux (m$^2$ s$^{-1}$) |
| $k_B$ | Boltzmann constant (J K$^{-1}$) |
| $k_m$ | thermal conductivity of membrane (W K$^{-1}$ m$^{-1}$) |
| $k_s$ | thermal conductivity of a solution (W K$^{-1}$ m$^{-1}$) |
| $l_{Du}$ | Dukhin length (m) |
| $L$ | nanopore length (m) |
| $L_r$ | reservoir length (m) |
| $L_i$ | coefficient given by Eqs. (D11)-(D15) (-) |
| $M_i$ | coefficient given by Eqs. (D16)-(D19) (-) |
| $M_{to}$ | thermoosmotic coefficient (m$^2$ s$^{-1}$) |
| $n_i$ | number concentration of ionic species $i$ (m$^{-3}$) |
| $N_i$ | coefficient given by Eqs. (D4)-(D7) (-) |
| $p$ | pressure (Pa) |
| $P_{max}$ | maximum power density (W m$^{-2}$) |
| $Pe$ | intrinsic ion Péclet number (-) |
| $Pe_i$ | Péclet number of ionic species $i$ (-) |
| $Pe_T$ | thermal Péclet number (-) |
| $r$ | radial coordinate (m) |
| $R_{ac}$ | access resistance (Ω) |
| $R_p$ | nanopore resistance (Ω) |
| $R_{res}$ | reservoir resistance (Ω) |
| $R_{tot}$ | total resistance (Ω) |
| $Re$ | Reynolds number (-) |
| $S_e$ | Seebeck coefficient (V K$^{-1}$) |
| $T$ | temperature (K) |
| $\Delta T$ | applied temperature difference (K) |
| $\Delta T_p$ | temperature difference between two nanopore ends (K) |
| $\boldsymbol{u}$ | fluid velocity (m s$^{-1}$) |
| $u_x$ | axial component of fluid velocity (m s$^{-1}$) |
| $u_{to}$ | average thermoosmotic velocity (m s$^{-1}$) |
| $x$ | axial coordinate (m) |
| $z_i$ | valence of ionic species $i$ (-) |

*Greek symbols*

| | |
|---|---|
| $\alpha$ | reduced Soret coefficient of ions (-) |
| $\alpha_i$ | reduced Soret coefficient of ionic species $i$ (-) |
| $\beta$ | adjustable parameter in Eq. (21) (-) |
| $\gamma$ | normalized difference in reduced Soret coefficients (-) |
| $\varepsilon_0$ | vacuum permittivity (F m$^{-1}$) |
| $\varepsilon_r$ | relative permittivity of solution (-) |
| $\kappa_\infty$ | reference Debye parameter (m$^{-1}$) |
| $\eta$ | dynamic viscosity of fluid (Pa s) |
| $\lambda_D$ | Debye length (m) |
| $\rho$ | density of solution (kg m$^{-3}$) |
| $\rho_e$ | free charge density (C m$^{-3}$) |
| $\sigma$ | electric conductivity (S m$^{-1}$) |
| $\Sigma$ | surface charge density (C m$^{-2}$) |
| $\chi$ | normalized difference in diffusion coefficients (-) |
| $\psi$ | double-layer potential (V) |
| $\Psi$ | dimensionless double-layer potential (-) |
| $\phi$ | electric potential (V) |
| $\Delta\phi$ | voltage (V) |

*Subscripts*

| | |
|---|---|
| ac | access |
| c | cold |
| h | hot |
| m | membrane |
| oc | open circuit |
| p | nanopore |
| res | reservoir |
| ref | reference |
| s | solution |
| sc | short circuit |
| to | thermoosmotic |
| v | virtual |
| ∞ | bulk |

# 1 Introduction

Thermo-osmotic response (TOR) [1-3] and thermoelectric response (TER) [4-7] generally occur simultaneously in a charged nanopore filled with an aqueous electrolyte solution when an axial temperature gradient is applied. They have diverse promising applications, such as low-grade heat energy conversion, charge separation, and desalination [8-10]. For promoting their practical applications, the major challenge confronting us is to enhance the two responses with an economical, nontoxic, and environment-friendly approach. To this end, it is necessary and imperative to unveil their physical mechanisms and further clarify the corresponding influencing factors, which are the cornerstone of the TOR and TER enhancement.

The TOR was first rationalized by Derjaguin and his colleagues [1, 11], who expressed the velocity far from the surface in terms of excess specific enthalpy. To date, most studies have focused on the TOR of pure water in nanotubes [12] or nanochannels [13, 14] and usually explained it as a Marangoni-like effect caused by the modification of interfacial tension [15, 16]. Moreover, some attempts have been made to study the TOR of aqueous electrolyte solutions in charge nanopores [5] or nanochannels [3, 17]. Semi-analytical models were developed for the slit nanochannels [3] or nanocapillaries [5] with sufficiently small height-to-length or radius-to-length ratios based on continuum theory along with the lubrication theory. These works clarified three fundamental origins of the TOR [3], namely, the chemiosmotic, thermal, and electroosmotic origins [17]. Specifically, the chemiosmotic flow is triggered by the osmotic pressure gradient due to the thermally induced ion concentration gradients, which is dictated by the Soret effect together with the ion imbalance inside the electric double layer (EDL) close to the charged nanopore wall. The thermal origin originates from the osmotic pressure gradient directly associated with the temperature gradient. The last TOR is propelled by the induced electric field. In addition, the pioneering experimental work [18] reported the microscopic observation of the thermoosmotic flow field for both untreated glass and Pluronic F-127-covered surfaces for the first time. Furthermore, molecular dynamics simulations were carried out together with the Onsager relation to investigate the microscopic picture of the TOR [2, 13]. These works help raise the awareness of the TOR in nanoconfined electrolyte solutions. Most recently, a giant enhancement in the TOR was reported in polyelectrolyte-brush-grafted nanochannels as compared with the brush-free nanochannels as long as the system is properly designed [19]. This initiates the way to boost the TOR in nanofluidic systems using polyelectrolyte-brush-grafted nanochannels.

For the TER, some attempts have also been made to uncover the secret of its physical mechanisms. For instance, in addition to well-known TER due to the difference in the Soret coefficients between cation and anion [20], Dietzel and Hardt discovered two categories of the thermoelectric mechanisms of electrolyte solutions in the charged nanochannels, namely, the thermally induced electric field due to the temperature-dependent electrophoretic effect [4] and the thermoosmotic streaming potential [3]. Zhang et al. [5] clarified the interplay of different TER mechanisms and put forward a synergy condition for the aforementioned thermoelectric effects to fully cooperate. Würger [21] clarified the difference in the TERs between open and closed systems and showed that for the zero-current steady state, the thermopower is independent of the ion mobilities. In addition, a new mechanism causing a giant TER was found using molecular dynamics simulation; that is, the excess specific enthalpy of water molecules together with the electro-osmotic mobility of solid-liquid interface could give rise to a TER and such a response would be enhanced by large slippage [22]. Recent numerical studies by Zhong and Huang [6, 23] showed that there is an optimal length-to-height ratio and optimal temperature difference maximizing the short-circuit current (SCC) and the open-circuit voltage (OCV) of the thermoelectricity in nanoconfined electrolyte solutions. These works motivate efforts to put the electrolyte-based TER to practical use.

However, there exist some ideal assumptions that deviate from the practical scenarios. First, the thermal conductivity of the nanofluidic membrane (i.e., the bulk material for fabricating nanopores), as one of the most important material parameters, was not explicitly taken into account in previous studies. To our knowledge, a thermal insulation boundary condition [6, 23] or a given temperature with a constant gradient [3, 4, 17] was frequently assumed at the solid-liquid interface in previous studies. The former assumption, equivalently, regards the thermal conductivity of the membrane as zero and is completely impractical; the latter should consume large amounts of heat energy to maintain in the practical application. Second, previous studies were in general based on the lubrication theory and neglected the effect of the reservoirs/the entrance effect [3, 5]. These assumptions would be invalid for finite-long nanopores and/or non-zero membrane thermal conductivity. Some investigations even further neglected the contributions of fluid flow or advective effect [4, 23], and thus the results can be inaccurate to some extent when the EDL thickness and the nanopore radius are comparable [5]. Third, the Soret equilibrium was frequently assumed in previous investigations [3-5, 17, 19], and thus the developed theory can only be valid for the mass closed system. However, this assumption does not hold anymore for the open system which is required for continuous device operation.

Therefore, previous theoretical and numerical studies did not provide a comprehensive physical picture of the TER and TOR of confined electrolytes, although they can partially explain some experimental observations. In this study, we aim to formulate a theoretic framework of the TER and TOR of nanoconfined electrolytes and further to have a comprehensive understanding of the TER and TOR. The theoretical framework discards the oversimplified assumptions and can overcome the limitations mentioned above. Accordingly, the theoretical framework could capture a more detailed physical picture of the TER/TOR and further guide the enhancement of the TER and/or the TOR, which is another goal of this paper. For such purposes, we systematically investigate the TER and the TOR in nanoconfined electrolyte solutions by developing both numerical and semi-analytical models. With the developed models, the influence of the thermal conductivity ratio, the surface charge density, and the membrane thickness on the TER and the TOR are thoroughly analyzed. Furthermore, some criteria are proposed for the enhancement of the TER/TOR.

## 2  Methods

*2.1 Problem description*

Figure 1a shows the investigated system which is comprised of a nanofluidic membrane (which refers to bulk materials for nanopore fabrication such as polyimide, polyethylene terephthalate, and silica) with a nanopore (whose radius and length are *a* and *L*, respectively) connected to two reservoirs at two ends. The nanopore, as a basic element of the nanofluidic membrane, is filled with an aqueous electrolyte solution and is subjected to an axial temperature difference, $\Delta T = T_\mathrm{h} - T_\mathrm{c}$, where $T_\mathrm{h}$ and $T_\mathrm{c}$ are temperatures of hot and cold sources, respectively. The EDL builds up in the region adjacent to the charged wall due to the redistribution of ions, and the EDL potential becomes inhomogeneous in the axial direction of the nanopore under non-isothermal conditions. With the application of the temperature difference, a current that originates from the transport of the ionic charges in the mobile part of the EDL (i.e., the electric diffusion layer beyond the shear plane [24]) emerges. Then, the accumulation of the ionic charges results in an electric field such that another current in an opposite direction to the aforementioned one. Meanwhile, the accompanying thermoosmotic flow takes place in the electrolyte solution. When the system reaches a steady state, the net current is zero; that is, the ionic fluxes vanish. The Seebeck coefficient [5], $S_\mathrm{e}$, and the thermoosmotic coefficient (TOC) [18], $M_\mathrm{to}$, are respectively two key parameters to describe the TER and TOR defined as

$$S_\mathrm{e} = -\frac{(\Delta\phi)_\mathrm{oc}}{\Delta T}, \quad M_\mathrm{to} = -\frac{u_\mathrm{to}}{\nabla T/T} \qquad (1)$$

where $(\Delta\phi)_{oc}$ is the induced electric potential difference under open-circuit conditions (that is, the OCV) and $u_{to}$ is the average thermo-osmotic velocity denoted by

$$u_{to} = \frac{1}{A}\int_\Omega dA \boldsymbol{e}_x \cdot \boldsymbol{u} \qquad (2)$$

with $\boldsymbol{u}=(u_x, u_r)$ being the velocity of the fluid, $A$ being the cross-section area of the nanopore, or reservoir ends, $\boldsymbol{e}_x$ being the unit vector pointing toward the $x$ direction, $\Omega$ denoting any cross-section.

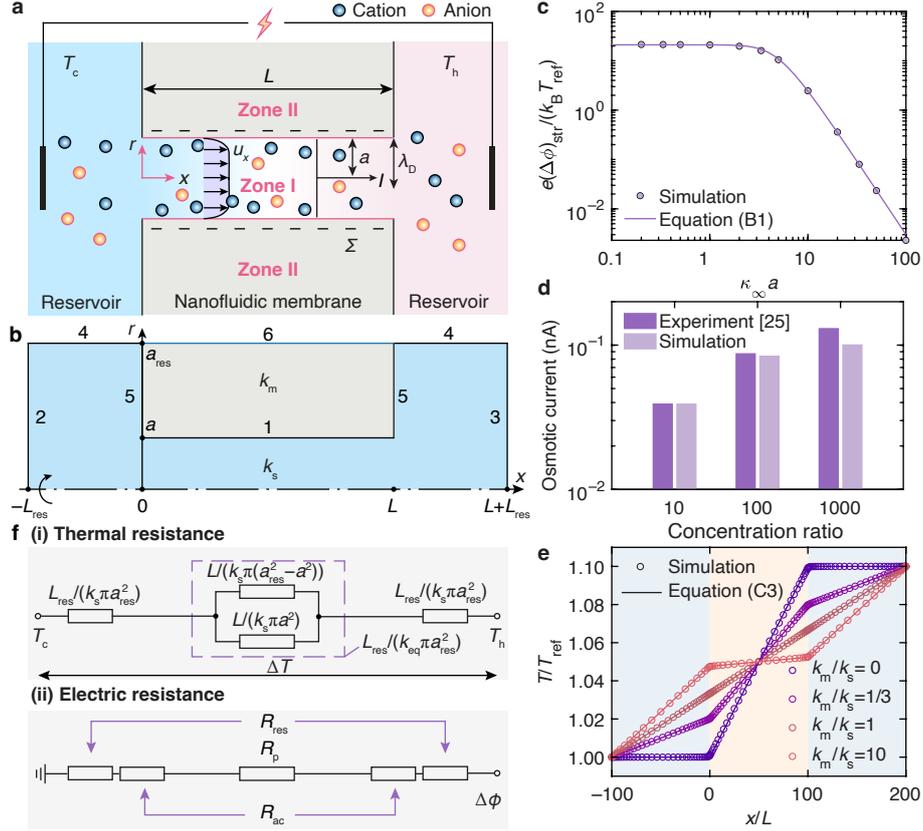

**Figure 1**: **Investigated system and model validation.** (**a**) Schematic of a charged nanopore of length $L$ and radius $a$ with two reservoirs at two ends. The cylindrical coordinate system is employed in this study with $r$ and $x$ respectively being the radial and axial directions. (**b**) Schematic of the two-dimensional axisymmetric computational domain. The length and radius of the reservoirs are labeled as $L_{res}$ and $a_{res}$, respectively. (**c**) Streaming potential $(\Delta\phi)_{str}$ normalized by $k_B T_{ref}/e$ as a function of dimensionless Debye parameter, $\kappa_\infty a$, for aqueous NaCl solutions with a dimensionless surface charge density of $\bar{\Sigma} = -10$, a dimensionless pressure difference of $\Delta\bar{P} = -1000$ and a dimensionless temperature difference of $\Delta\bar{T} = 0$. Symbols stand for the numerical results and lines stand for semi-analytical results obtained from Eq. (B1) (see Appendix B). (**d**) Comparison between the numerical and experimental results of the salinity-gradient induced osmotic current of the aqueous KCl solution in a boron nitride nanotube of radius $a = 40$ nm and length $L = 1024$ nm at pH 5.5. The experimental data are taken from Ref. [25] with the surface charge density being determined as -60 mC m$^{-2}$ [26]. (**e**) Dimensionless temperature profile along the centerline of the nanopore and the reservoirs ($r=0$) for $\Delta\bar{T} = 0.1$. Symbols stand for the numerical results and lines stand for semi-analytical results obtained from Eq. (C3). (**f**) Schematic of thermal resistance (i) and the internal electric resistance (ii).

## 2.2 Governing equations and boundary conditions

The investigated system considered in this study (Figure 1a) can be described by the energy equation together with the extended Nernst-Planck-Poisson equations and the Navier-Stokes equation, which read

In *zone I*:

$$\rho c_p \boldsymbol{u} \cdot \nabla T = \nabla \cdot (k_s \nabla T) + \Phi_\eta + \Phi_\phi \tag{3}$$

$$\nabla \cdot \boldsymbol{J}_i = 0 \text{ with } \boldsymbol{J}_i = \boldsymbol{u} n_i - D_i \left( \nabla n_i + 2\alpha_i n_i \frac{\nabla T}{T} + \frac{z_i e n_i}{k_B T} \nabla \phi \right) \tag{4}$$

$$\nabla \cdot (\varepsilon_0 \varepsilon_r \nabla \phi) = -\rho_e \tag{5}$$

$$\nabla \cdot \boldsymbol{u} = 0, \rho \boldsymbol{u} \cdot \nabla \boldsymbol{u} = -\nabla p + \nabla \cdot \{\eta[\nabla \boldsymbol{u} + (\nabla \boldsymbol{u})^T\} - \rho_e \nabla \phi - \frac{\varepsilon_0}{2} |-\nabla \phi|^2 \nabla \varepsilon_r \tag{6}$$

In *zone II*:

$$\nabla \cdot (k_m \nabla T) = 0 \tag{7}$$

where $T$ is the absolute temperature, $\phi$ is the electric potential, $p$ is the pressure; $\boldsymbol{J}_i, n_i, D_i, \alpha_i$ and $z_i$ are the flux, the number concentration, the diffusion coefficient, the reduced Soret coefficient [20] and the valence of the ionic species $i$, respectively; $\rho, c_p, k_s, \varepsilon_r, \eta$ are the density, the heat capacity at constant pressure, the thermal conductivity, the relative permittivity and the dynamic viscosity of the electrolyte solution, respectively; $k_m$ is the thermal conductivity of the membrane; $\varepsilon_0$, $e$ and $k_B$ are the vacuum permittivity, the elementary charge and the Boltzmann constant, respectively; $\rho_e = e \sum_i z_i n_i$ is the free charge density in zone I; $\Phi_\eta = \tau : \nabla \boldsymbol{u}$ is the viscous dissipation [27] with $\tau = \eta[\nabla \boldsymbol{u} + (\nabla \boldsymbol{u})^T]$ being the viscous stress tensor and $\Phi_\phi = \sigma |\nabla \phi|^2$ is the Joule heating with $\sigma$ being the local electric conductivity. These two heat source terms have been shown to be negligible [5].

To simplify the analysis, we nondimensionalize the gradient operator, the temperature, the electric potential, the ion number concentration, the velocity, and the pressure by division with the following reference quantities respectively:

$$[\nabla] = \frac{1}{a}, [T] = T_{\text{ref}}, [\phi] = \frac{k_B T_{\text{ref}}}{e} \equiv \phi_{\text{ref}}, [n_i] = \frac{1}{2} \sum_i z_i^2 n_{i,\infty}, [\boldsymbol{u}] = \frac{\varepsilon_0 \varepsilon_{r,\text{ref}}[\phi]}{\eta_{\text{ref}}} \frac{[\phi]}{a} \equiv u_{\text{ref}}, [p] = \frac{\eta_{\text{ref}} u_{\text{ref}}}{a} \equiv p_{\text{ref}} \tag{8}$$

where the subscript ref denotes the physical properties taken at the reference temperature $T_{\text{ref}} = 298.15$ K and subscript $n_{i,\infty}$ refers to the bulk ion number concentration. With these reference variables, Eqs. (3) to (7) can be transformed to

In *zone I*:

$$Pe_T \bar{\boldsymbol{u}} \cdot \bar{\nabla} \bar{T} = \bar{\nabla}^2 \bar{T} \tag{9}$$

$$\overline{\nabla} \cdot \overline{\boldsymbol{J}}_i = 0 \text{ with } \overline{\boldsymbol{J}}_i = \overline{\boldsymbol{u}}\bar{n}_i - \frac{\overline{D}_i}{Pe_i}\left(\overline{\nabla}\bar{n}_i + \frac{2\alpha_i\bar{n}_i}{\overline{T}}\overline{\nabla}\overline{T} + \frac{z_i\bar{n}_i}{\overline{T}}\overline{\nabla}\bar{\phi}\right) \quad (10)$$

$$\overline{\nabla} \cdot (\bar{\varepsilon}_r \overline{\nabla}\bar{\phi}) = -\frac{(\kappa_\infty a)^2}{2}\sum_i z_i \bar{n}_i \quad (11)$$

$$\overline{\nabla} \cdot \overline{\boldsymbol{u}} = 0, \; Re\overline{\boldsymbol{u}} \cdot \overline{\nabla}\overline{\boldsymbol{u}} = -\overline{\nabla}\bar{p} + \overline{\nabla} \cdot \{\bar{\eta}[\overline{\nabla}\overline{\boldsymbol{u}} + (\overline{\nabla}\overline{\boldsymbol{u}})^{\mathrm{T}}]\} - \frac{(\kappa_\infty a)^2}{2}\sum_i z_i \bar{n}_i \overline{\nabla}\bar{\phi} - \frac{1}{2}|\overline{\nabla}\bar{\phi}|^2 \overline{\nabla}\bar{\varepsilon}_\mathrm{r} \quad (12)$$

In *zone II*:

$$\overline{\nabla} \cdot [(k_\mathrm{m}/k_\mathrm{s})\overline{\nabla}\overline{T}] = 0 \quad (13)$$

where

$$\kappa_\infty a = a\sqrt{\sum_i \frac{e^2 z_i^2 n_{i,\infty}}{\varepsilon_0 \varepsilon_{\mathrm{r,ref}} k_\mathrm{B} T_\mathrm{ref}}} \equiv \frac{a}{\lambda_\mathrm{D}}, \; Pe_T = \frac{u_\mathrm{ref}a}{k_\mathrm{s}/\rho c_p}, \; Pe_i = \frac{u_\mathrm{ref}a}{D_{i,\mathrm{ref}}}, \; Re = \frac{\rho u_\mathrm{ref}a}{\eta_\mathrm{ref}} \quad (14)$$

are the ratio of the nanopore radius to the Debye length $\lambda_\mathrm{D}$ (which is termed the dimensionless Debye parameter or the double-layer thickness parameter [28]), the reference thermal Péclet number, the reference Péclet number, and the reference Reynolds number, respectively. In addition, $\bar{\eta} = \eta/\eta_\mathrm{ref}$, $\bar{\varepsilon}_\mathrm{r} = \varepsilon_\mathrm{r}/\varepsilon_\mathrm{r,ref}$ and $\overline{D}_i = D_i/D_{i,\mathrm{ref}}$ are functions of the temperature (see Appendix A) and the bars on notations designate their dimensionless counterparts. Furthermore, the dimensionless electric current is evaluated by

$$\bar{I} = \frac{I}{[I]} = \frac{1}{A}\int_\Omega \mathrm{d}A \sum_i \hat{\boldsymbol{e}}_x \cdot z_i \overline{\boldsymbol{J}}_i \quad (15)$$

where $[I]=\pi a^2 e[\boldsymbol{u}][n_i]$ is the reference electric current.

Figure 1b shows the two-dimensional axisymmetric computational domain and all boundaries are labeled by integers. The boundary conditions corresponding to Eqs. (9) to (13) are summarized in Table 1. A few remarks on the boundary conditions are in order. First, on the nanopore wall (boundary 1), a constant surface charge density rather than a constant surface potential was used by noting that on the one hand, $\phi$ is the superposition of the EDL potential, $\psi$, and the induced potential, $\phi_\mathrm{v}$, and the two components are inseparable with the latter being yet to determined; on the other hand, the surface charge density is nearly constant at a fixed pH, independent of bulk concentrations [25]. Although depending on specific physio-chemical reactions, the temperature dependence of the surface charge density is neglected for a relatively small range of temperature variation considered in this study ($\leq$ 30 K), especially for silica surfaces [4]. Second, the TER is unknown beforehand under open-circuit conditions; thus the potential of the leftmost reservoir boundary (boundary 2) can be designated to be zero, whereas that of the rightmost reservoir boundary (boundary 3) is obtained by the zero-current constraint [5, 29] or parametric sweep studies with interpolation [4]. Third, the no-slip boundaries are considered in the nanopore wall and membrane surface (boundaries 1 and 4). Finally, the constant pressure at the leftmost and the rightmost of the reservoirs (boundaries 2 and 3) can be approximately achieved by an open boundary [10].

**Table 1: Boundary conditions for Eqs. (9) to (13) in the present study.** Here, $\hat{n}$ is the outward unit normal vector, the surface charge density is nondimensionalized by $\varepsilon_0 \varepsilon_{r,\mathrm{ref}} k_B T_{\mathrm{ref}}/(ea)$.

| Boundaries | Eq. (9) or Eq. (13) | Eq. (10) | Eq. (11) | Eq. (12) |
|---|---|---|---|---|
| 1 | Heat flux conversion $\hat{n}\cdot\overline{\nabla}\overline{T}\vert_\mathrm{s} = \hat{n}\cdot(k_\mathrm{m}/k_\mathrm{s})\overline{\nabla}\overline{T}\vert_\mathrm{m}$ | No flux $\hat{n}\cdot\bar{J}_i = 0$ | Constant charge $-\hat{n}\cdot\bar{\varepsilon}_r\overline{\nabla}\bar{\phi} = \bar{\Sigma}$ | No slip $\bar{u} = \vec{0}$ |
| 2 | Constant temperature $\bar{T} = T_\mathrm{c}/T_\mathrm{ref}$ | Bulk concentration $\bar{n}_i = 1$ | Ground $\bar{\phi} = 0$ | Constant pressure $\bar{p} = 0$ |
| 3 | Constant temperature $\bar{T} = T_\mathrm{h}/T_\mathrm{ref}$ | Bulk concentration $\bar{n}_i = 1$ | Voltage bias $\bar{\phi} = \Delta\bar{\phi}$ | Constant pressure $\bar{p} = 0$ |
| 4 | Heat insulation $\hat{n}\cdot\overline{\nabla}\overline{T} = 0$ | No flux $\hat{n}\cdot\bar{J}_i = 0$ | No charge $\hat{n}\cdot\overline{\nabla}\bar{\phi} = 0$ | Slip |
| 5 | Heat flux conversion $\hat{n}\cdot\overline{\nabla}\overline{T}\vert_\mathrm{s} = \hat{n}\cdot(k_\mathrm{m}/k_\mathrm{s})\overline{\nabla}\overline{T}\vert_\mathrm{m}$ | No flux $\hat{n}\cdot\bar{J}_i = 0$ | No charge $\hat{n}\cdot\overline{\nabla}\bar{\phi} = 0$ | No slip $\bar{u} = \vec{0}$ |
| 6 | Heat insulation $\hat{n}\cdot\overline{\nabla}\overline{T} = 0$ | – | – | – |

## 2.3  Numerical approach and validation

In this work, we carried out numerical simulations using the finite element software COMSOL Multiphysics. Five built-in modules were used, that is an *Electrostatic* module for the electric potential, a *Laminar Flow* module for the flow field, a *Heat Transfer in Solids and Liquids* module for the temperature, and two *General PDE* modules for the ion concentrations. Notably, the term $(1/\bar{r})\partial_{\bar{r}}(\cdot)$ should be complemented as a source term in the *General PDE* modules because the built-in divergence $\nabla^2(\cdot)$ in such modules does not take into account the effect of the curved coordinate system. In addition, the zero-current constraints can be achieved by a *Global ODE* module self-consistently [5, 29].

**Table 2: Parameter values used in this study.** The diffusion coefficient and reduced Soret coefficient of Na$^+$ and Cl$^-$ are taken from Ref. [28] and Ref. [30], respectively. In addition, $I_0 = \pi a^2 F[u]C_0$ is used to rescale the current for different $\kappa_\mathrm{ref}a$ with $C_0$=1 mM and $F$ being the Faraday constant; $Pe = u_\mathrm{ref}a/D_\mathrm{ref} = Pe_+Pe_-/(Pe_++Pe_-)$ with $D_\mathrm{ref} = (D_{+,\mathrm{ref}}+D_{-,\mathrm{ref}})/2$.

| Reference properties | | Reference quantities | | Dimensionless parameters | |
|---|---|---|---|---|---|
| $D_{+,\mathrm{ref}}$ (×10$^{-9}$ m$^2$ s$^{-1}$) | 1.334 | $a$ (nm) | 10 | $Pe_+$ | 0.38645 |
| $D_{-,\mathrm{ref}}$ (×10$^{-9}$ m$^2$ s$^{-1}$) | 2.032 | $T_\mathrm{ref}$ (K) | 298.15 | $Pe_-$ | 0.2537 |
| $\eta_\mathrm{ref}$ (×10$^{-4}$ Pa s) | 8.9 | $u_\mathrm{ref}$ (m s$^{-2}$) | 0.051552 | $Pe$ | 0.30631 |
| $\varepsilon_\mathrm{ref}$ | 78.408 | $\Sigma_\mathrm{ref}$ (mC m$^{-2}$) | 17.858 | $Pe_T$ | 0.0035069 |
| $\rho_\mathrm{ref}$ (kg m$^{-3}$) | 997 | $P_\mathrm{ref}$ (Pa) | 4588.1 | $Re$ | 5.775×10$^{-4}$ |
| $c_{p,\mathrm{ref}}$ (kJ kg$^{-1}$ K$^{-1}$) | 4.1 | $\phi_\mathrm{ref}$ (mV) | 25.693 | $\alpha_+$ | 0.7 |
| $k_{s,\mathrm{ref}}$ (W m$^{-1}$ K$^{-1}$) | 0.6 | $I_0$ (pA) | 1.5626 | $\alpha_-$ | 0.1 |

As depicted in Figure 1b, unless otherwise stated, the capillary length was chosen as $L=100a$ with the nanopore radius being set to $a$=10 nm, while the radius and length of reservoirs were set to $a_\mathrm{res}$=100$a$ and $L_\mathrm{res}$=100$a$, respectively. The nanopore was chosen as being 100×1000 cells with an element ratio of (at least) 10 in the radial direction to refine the near-wall region and capture the EDL characteristics; the membrane was chosen as being 1000×1000 cells with an element ratio of (at least) 100 in the radial direction (symmetrically distributed); the reservoirs adopted free triangular mesh whose element size is predefined as extremely fine. The mesh independence study has confirmed that the present mesh can achieve robust results.

Figure 1c shows the comparison of the streaming potential between numerical and semi-analytical results for various Debye parameters. The numerical results are in excellent agreement with the semi-analytical results. In addition, Figure 1d indicates that the numerical model can achieve satisfactory predictions for the experimental data of the osmotic current of the aqueous KCl solution in a boron nitride nanotube with radius $a = 40$ nm and length $L = 1024$ nm at pH 5.5 under the isothermal condition [25]. Furthermore, the numerical results manifest that when $k_m/k_s=0$ (in the calculation, $k_m$ is set to be a small value such as $10^{-6}$), the temperature linearly varies along the axial direction of the nanopore and there is no temperature difference in the reservoirs (Figure 1e), which agrees well with the analytical solutions obtained by the order of magnitude analysis [5]. Also, the temperature distribution in the system is well captured by the thermal resistance analysis regardless of $k_m/k_s$ values as $L/a=100$ (Figure 1e). Therefore, our numerical model can provide a reasonable and highly robust estimation of the TOR and the TER. In this study, the reference quantities and dimensionless parameters are estimated and listed in Table 2. In addition, the temperature conditions are set to $\bar{T}_c = 1$ and $\Delta\bar{T} = 0.1$.

*2.4    Semi-analytical model*

Under the condition of $(a/L)^2 \ll 1$, the lubrication approximation was employed to develop a semi-analytical model for the TER and the TOR of a confined electrolyte solution in the Soret equilibrium as $k_m/k_s = 0$ [3, 5]. Here, an extended semi-analytical model would be developed by discarding the assumptions of the Soret equilibrium and the zero $k_m/k_s$. To begin with, with the help of the thermal resistance network (Figure 1f, top), the temperature distribution is found to be described by Eq. (C3) (see Appendix C for details). Later, the potential is assumed to be the superposition of the EDL potential, $\psi$, and the induced potential, $\phi_v$ [31, 32]. The former is governed by the well-known Poisson-Boltzmann equation (Eqs. (D8) and (D9)). As detailed in Appendix D, the dimensionless average velocity, $\bar{u}_{to}$, the dimensionless ion flux, $\bar{J}_{ion} = 2\int_0^1(\bar{J}_{+,x} + \bar{J}_{-,x})\bar{r}d\bar{r}$, and the dimensionless current, $\bar{I} = 2\int_0^1(\bar{J}_{+,x} - \bar{J}_{-,x})\bar{r}d\bar{r}$, in the nanopore are derived as

$$\begin{bmatrix} \bar{u}_{\text{to}} \\ \bar{J}_{\text{ion}} \\ \bar{I} \end{bmatrix} = \begin{bmatrix} L_1 & L_2 & L_3 & L_4 \\ M_1 & M_2 & M_3 & M_4 \\ N_1 & N_2 & N_3 & N_4 \end{bmatrix} \begin{bmatrix} -\partial_{\bar{x}}\bar{p}_{\text{v}} \\ -\partial_{\bar{x}}\bar{n}_{\text{v}} \\ -\partial_{\bar{x}}\bar{T} \\ -\partial_{\bar{x}}\bar{\phi}_{\text{v}} \end{bmatrix} \qquad (16)$$

where $\bar{p}_{\text{v}}(x) = \bar{p} - (\kappa_\infty a)^2 \bar{n}_{\text{v}} \bar{T}[\cosh(e\psi/k_{\text{B}}T) - 1]$ is known as the virtual hydrostatic pressure [5] and $n_{\text{v}}(x) = (\bar{n}_+\bar{n}_-)^{1/2}$ is the virtual number concentration of each ionic species [5, 31]. In addition, the coefficients, $L_i$, $M_i$ and $N_i$, are given in Appendix D.

Here, the heat flux is not formulated since its inclusion would be quite challenging and can be left for future investigation. This is because, in contrast to other fluxes that are only transferred through the solution in the nanopore, the heat flux is transferred in both the nanopore and the membrane. For simplicity, in this study, the temperature distribution is determined by analyzing the thermal resistance network. Clearly, replacing the present thermodynamic forces, $-\partial_{\bar{x}}\bar{p}_{\text{v}}$ and $-\partial_{\bar{x}}\bar{n}_{\text{v}}$ with $-[2/(\kappa_\infty a)^2]\partial_{\bar{x}}\bar{p}_{\text{t}} = -[2/(\kappa_\infty a)^2]\partial_{\bar{x}}\bar{p}_{\text{v}} + 2T\partial_{\bar{x}}\bar{n}_{\text{v}} + 2\bar{n}_{\text{v}}\partial_{\bar{x}}\bar{T}$ and $-\bar{T}\partial_{\bar{x}} \ln \bar{n}_{\text{v}}$ (in order)[2], one can readily check that the Onsager relation is satisfied. In addition, in the absence of a temperature gradient, Eq. (16) can recover the classic space-charge model under isothermal conditions [33].

In the reservoirs, Eq. (16) can be simplified as

$$\begin{bmatrix} \bar{u}_{\text{to}} \\ \bar{J}_{\text{ion}} \\ \bar{I} \end{bmatrix} = \frac{a_{\text{res}}^2}{a^2} \frac{2\bar{D}}{Pe} \begin{bmatrix} \frac{Pe}{16\bar{\eta}\bar{D}} & 0 & 0 & 0 \\ \frac{\bar{n}_{\text{v}}Pe}{8\bar{\eta}\bar{D}} & 1 & 2\alpha\frac{\bar{n}_{\text{v}}}{\bar{T}} & \chi\frac{\bar{n}_{\text{v}}}{\bar{T}} \\ 0 & \chi & 2\gamma\alpha\frac{\bar{n}_{\text{v}}}{\bar{T}} & \frac{\bar{n}_{\text{v}}}{\bar{T}} \end{bmatrix} \begin{bmatrix} -\partial_{\bar{x}}\bar{p}_{\text{v}} \\ -\partial_{\bar{x}}\bar{n}_{\text{v}} \\ -\partial_{\bar{x}}\bar{T} \\ -\partial_{\bar{x}}\bar{\phi}_{\text{v}} \end{bmatrix} \qquad (17)$$

where $Pe = u_{\text{ref}}a/D_{\text{ref}}$ is known as the intrinsic Péclet number [34] with $D_{\text{ref}} = (D_{+,\text{ref}} + D_{-,\text{ref}})/2$, $\bar{D} = \sum_i D_i / \sum_i D_{i,\text{ref}}$ the average ion diffusion coefficient, $\alpha = \sum_i \alpha_i D_i / \sum_i D_i$ the reduced ion Soret coefficient, $\chi = (D_+ - D_-)/(D_+ + D_-)$ the normalized difference in the diffusion coefficients of cation and anion, $\gamma = (\alpha_+ D_+ - \alpha_- D_-)/\sum_i \alpha_i D_i$ the normalized difference in the diffusivity-modified Soret coefficients between cation and anion.

Currently, there are six unknowns in each part of the fluid zone (i.e., nanopore or reservoirs), namely, three fluxes ($\bar{u}_{\text{to}}$, $\bar{J}_{\text{ion}}$ and $\bar{I}$) and three virtual quantities ($\bar{p}_{\text{v}}$, $\bar{n}_{\text{v}}$ and $\bar{\phi}_{\text{v}}$), but only three equations. Therefore, three equations need to be added to close Eq. (16) or (17): $\partial_{\bar{x}}\varphi=0$ or $\varphi$=const, where $\varphi=\bar{u}_{\text{to}}$, $\bar{J}_{\text{ion}}$ or $\bar{I}$. For solving the above equations, proper boundary conditions are complemented as follows: $\bar{p}_{\text{v}}(-L_{\text{r}}/a) = \bar{p}_{\text{v}}(L/a + L_{\text{res}}/a) = 0$, $\bar{n}_{\text{v}}(-L_{\text{res}}/a) = \bar{n}_{\text{v}}(L/a + L_{\text{res}}/a) = 0$, $\bar{\phi}_{\text{v}}(-L_{\text{res}}/a) = 0$ and $\bar{\phi}_{\text{v}}=(L/a + L_{\text{res}}/a)=\Delta\bar{\phi}_{\text{v}}$. Furthermore, the three virtual quantities are assumed to be continuous at the nanopore-reservoir interfaces. Subsequently, the numerical implementation of the semi-analytical model is detailed in Appendix D. Finally, the SCC can be evaluated by Ohm's law as

$$I_{sc} = \frac{[\phi_v(L+L_{res})-\phi_v(-L_{res})]_{oc}}{R_{tot}} = \frac{(\Delta\phi_v)_{oc}}{R_{tot}} \quad (18)$$

where $R_{tot}$ is the total internal electrical resistance of the investigated system and will be analyzed in the following section.

---

[2] Typos on the replaced thermodynamic force in the published paper has been correted here, this correlation does not affect the results and findings in this paper.

## 2.5 Resistance analysis

The equivalent electric circuit is shown in the bottom panel of Figure 1f, and the total resistance of the investigated system ($R_{tot}$) is the superposition of the nanopore resistance, $R_p$, the *access resistance* due to nanopore-reservoir interface, $R_{ac}$ [35] and the reservoir resistance, $R_{res}$. Among them, the reservoir resistance is given as

$$R_{res} = \frac{1}{\sigma_{ref}} \frac{a}{\pi a_{res}^2} \left[ \int_{-\frac{L_{res}}{a}}^{0} \frac{\bar{T}}{\bar{D}\bar{n}_v} d\bar{x} + \int_{\frac{L}{a}+\frac{L_{res}}{a}}^{\frac{L}{a}+\frac{2L_{res}}{a}} \frac{\bar{T}}{\bar{D}\bar{n}_v} d\bar{x} \right] \approx \frac{2}{\sigma_{ref}} \frac{L_r}{\pi a_{res}^2} \frac{\bar{T}_{avg}}{\bar{D}(\bar{T}_{avg})} \quad (19)$$

where $\sigma_{ref} = 2e^2 n_\infty D_{ref}/(k_B T_{ref})$ is the reference electric conductivity, $\bar{T}_{avg}$ is the average temperature. Neglecting the contribution of the thermo-osmosis, we can approximate the nanopore resistance as

$$R_p = \frac{1}{\sigma_{ref}} \frac{1}{\pi a} \frac{2}{Pe} \int_0^{L/a} \frac{d\bar{x}}{N_4} \approx \frac{1}{\sigma_{ref}} \frac{L}{\pi a^2} \frac{2}{Pe} \frac{1}{N_4(L/2a)} \quad (20)$$

The access resistance is derived as

$$R_{ac} = \frac{1}{\sigma_{ref}} \left[ \frac{1}{4a+\beta l_{Du,\bar{x}=0}} + \frac{1}{4a+\beta l_{Du,\bar{x}=L/a}} \right] \approx \frac{1}{\sigma_{ref}} \frac{2}{4a+\beta l_{Du,\bar{x}=L/2a}} \quad (21)$$

where $\beta$ is an adjustable parameter, $l_{Du}$ is the Dukhin length measuring the relative importance of the surface conductance and the bulk conductance and is approximated by

$$l_{Du,\bar{x}} \approx \frac{a}{2} \left\{ 2 \int_0^1 \left[ \cosh\left(\frac{e\psi}{k_B T}\right) - \chi \sinh\left(\frac{e\psi}{k_B T}\right) \right] \bar{r} d\bar{r} - 1 \right\} \quad (22)$$

It is worth mentioning that the reservoir resistance is frequently neglected in a single nanopore configuration in the literature [25, 36-38]. In the absence of temperature difference and neglecting the contribution of reservoir resistance, the total resistance

$$R_{tot} = R_{res} + R_p + R_{ac} \quad (23)$$

recovers the literature expression [37]. The adjustable parameter $\beta$ can be determined by both experiment and numerical calculation and is obtained by minimizing the sum of square error based on numerical results here.

Clearly, with the temperature distribution given in Eq. (C3), the reservoir resistance can be directly determined by Eq. (19), and the nanopore resistance can be evaluated by Eq. (20) with the knowledge of $N_4$ obtained from Algorithm 1 (see Appendix D), and the access resistance is evaluated by Eq. (21) with the determined value of $\beta$ and the double-layer potential obtained from Algorithm 1.

## 3 Results and discussion

### 3.1 Thermoelectric response and its behavior

Generally, the thermal conductivity of the charged membrane is a finite value (see Table E1), resulting in the TOR and TER that differ from the ideal cases of $k_\mathrm{m}/k_\mathrm{s} = 0$ and $k_\mathrm{m}/k_\mathrm{s} = \infty$,. As $k_\mathrm{m}/k_\mathrm{s} = 0$, the nanopore wall is equivalent to the heat insulation surface and thus the effective temperature difference across the nanopore is identical to the applied temperature difference [39] (Figure 1d). For the opposite case of $k_\mathrm{m}/k_\mathrm{s} = \infty$, the heat flux would be directly transferred through the membrane rather than the aqueous electrolyte solution, and thus the solution in the nanopore becomes isothermal (Figure 1d and Eq. (C2)). For a finite value of $k_\mathrm{m}/k_\mathrm{s}$, the temperature distribution falls between these two limiting cases. Obviously, the difference in the temperature distribution could have a substantial effect on both the TOR and the TER. Despite this, it is observed from Figure 2a that the voltage-current characteristics are linear (i.e., in the Ohm region) and the zero-current constraints can well predict the OCV due to the consistency between the results obtained from the zero-current constraints (solid symbols, Figure 2a (i)) and those obtained by interpolating the voltage-current relation (dashed lines). This indicates there is no need to present inefficient parametric sweep studies to determine the OCV for $k_\mathrm{m}/k_\mathrm{s} = 0$. However, this strategy fails to evaluate the OCV for a finite thermal conductivity ratio.

We first consider the characteristics of the SCC. With the reference current defined in Section 2.2, the dimensionless SCCs for different $\kappa_\infty a$ could not be compared with each other directly since the reference current, $[I]$, is dependent on $n_\infty$ and thus $\kappa_\infty a$. Therefore, for comparison, the current is rescaled by $I_0=\pi a^2 F[u]C_0$ (whose value is given in Table 2) with $C_0$=1 mM and $F$ being the Faraday constant. Clearly, the SCC arrives at its maximum value at around $\kappa_\infty a$ ~3 and decreases if $\kappa_\infty a$ deviates from ~3 (Figure 2b). This qualitative behavior is in analogy to that of the streaming current [40] and has also been experimentally confirmed in conical nanochannels [41]. The increase in the access resistance with decreasing $\kappa_\infty a$ is responsible for this behavior by noting that Eq. (18) can well reproduce the numerical results (Figure 2d and Figure 3). When the access resistance is absent, Eq. (18) would fail to predict the resistance and the SCC accurately. For instance, for $\kappa_\infty a$ = 1, when the access resistance is absent, the resistance would decay to ~1/3 of its actual value (Figure 3) and accordingly the calculated

SCC would be ~3 times of its numerical result, confirming the existence of the access resistance due to the entrance effect. Such access resistance is essentially caused by the perturbation of the electric field lines and the bending of the current streamlines near nanopore mouths [37].

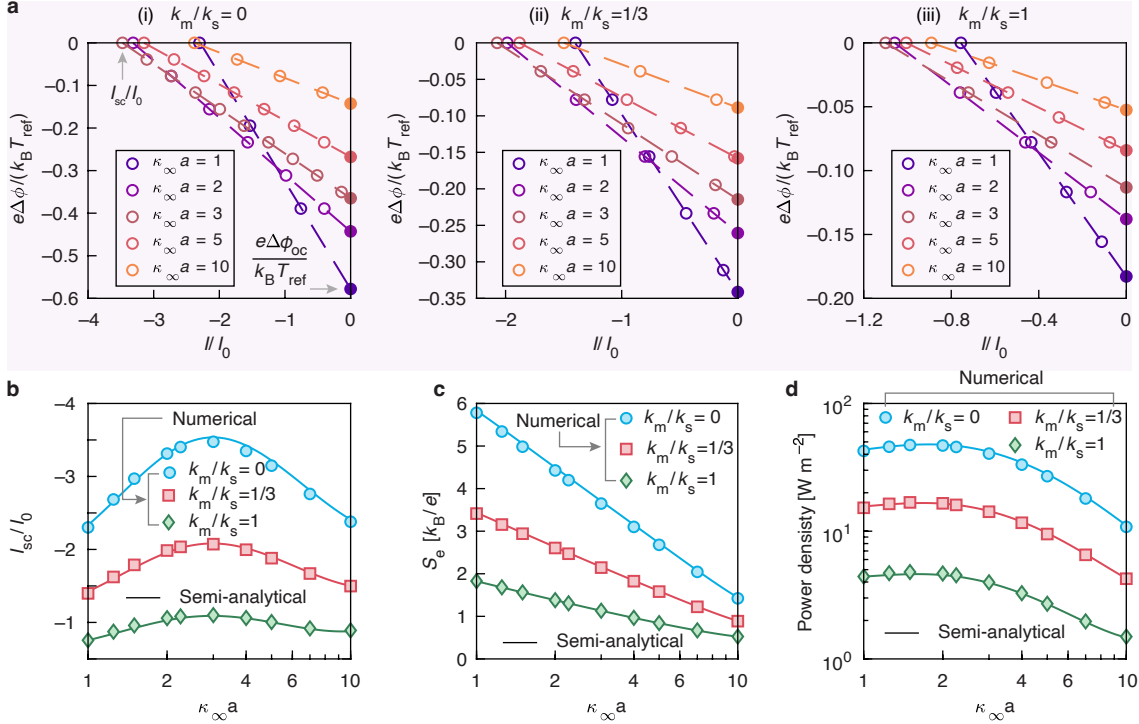

**Figure 2**: **Thermoelectric response**. (**a**) Voltage-current characteristics for various Debye parameters, $\kappa_\infty a$ at (i) $k_m/k_s =0$ and (ii) $k_m/k_s =1/3$ and (iii) $k_m/k_s =1$. The voltage, $\Delta\phi$, and the current, $I$, are normalized by $[\phi] = k_B T_{ref}/e$ and $I_0$, respectively. Dashed lines are linear fits, open symbols are obtained from parametric studies, and solid symbols are obtained from zero-current constraint (i) or interpolation (ii, iii). (**b**) Short circuit current (SCC) $I_{sc}$ normalized by $I_0$ as a function of the Debye parameter, $\kappa_\infty a$, for varying thermal conductivity ratios, $k_m/k_s$. Symbols stand for numerical results and lines are calculated by Eq. (18) with $R_{tot}$ being evaluated by Eq. (23), see Figure 3. (**c**) Seebeck coefficient $S_e$ normalized by $k_B/e$ as a function of dimensionless Debye parameter, $\kappa_\infty a$, for various thermal conductivity ratios, $k_m/k_s$. Symbols stand for numerical results and solid lines stand for results obtained from the semi-analytical model. (**d**) Maximum power density as a function of Debye parameter, $\kappa_\infty a$, for various thermal conductivity ratios. The surface charge density and the temperature difference were set to $\bar{\Sigma}$=50 and $\Delta\bar{T} = 0.1$, respectively.

We next focus on the Seebeck coefficient. Significantly, the numerical results of the Seebeck coefficient, $S_e$, are in excellent agreement with the semi-analytical results (Figure 2c), verifying the semi-analytical model again. It is observed that the Seebeck coefficient decreases with the Debye parameter, $\kappa_\infty a$, regardless of $k_m/k_s$ values. This behavior has been numerically demonstrated for the nanopores at $k_m/k_s$ =0 [4, 5] and is also in agreement with the experimental observations of the TER of electrolyte solutions in charged membrane systems [42, 43]. In addition, with increasing $k_m/k_s$, the Seebeck coefficient decreases because the effective temperature difference across the nanopore decreases with $k_m/k_s$ (Figure 1d and Figure C1).

Due to the linear voltage-current characteristics (Figure 2a), the maximum power density can be estimated by $P_{max} = I_{sc}\Delta V/(4\pi a^2)$. It is found that as $\kappa_\infty a \approx 1.5$, the maximum power density reaches its peak value and decreases when $\kappa_\infty a$ deviates from 1.5 (Figure 2d). For the present computational parameter values, the maximum power densities corresponding to $k_m/k_s$ =0, 1/3 and 1 are estimated to be 47.3 W m$^{-2}$, 16.8 W m$^{-2}$, and 4.8 W m$^{-2}$, respectively. For the ionic thermoelectric devices, it would be better to characterize their performance by the normalized maximum power density defined by $P_{max}A^{-1}\Delta T^{-2}$ [44], where $A$ is the cross-sectional area. The normalized maximum power densities are then estimated to be 53.2 mW K$^{-2}$ m$^{-2}$, 18.9 mW K$^{-2}$ m$^{-2}$, and 5.4 mW K$^{-2}$ m$^{-2}$, respectively. These values are at least an order of magnitude higher than those for thermoelectric supercapacitors and thermocells [44]. It is worth mentioning that the power densities above are defined based on the cross-sectional area of the nanopore and should be defined in terms of the membrane area in practical applications [45]. With a typical porosity of 0.3 [38], the estimated maximum power density corresponding to $k_m/k_s$ =0, 1/3, and 1 can be translated to be 14.2 W m$^{-2}$, 5 W m$^{-2}$ and 1.4 W m$^{-2}$, respectively. These values in general exceed most of the proof-of-concept low-grade heat energy generators for the same temperature difference [46-49].

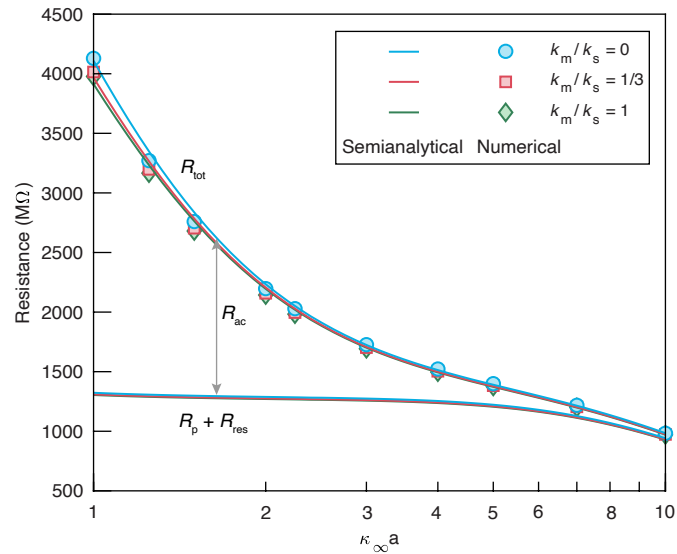

**Figure 3**: **Electric resistance of the investigated system**. Symbols stand for numerical results; solid lines stand for semi-analytical solutions obtained from Eq. (23) with $\beta = 0.036$, 0.041, and 0.043 corresponding to the thermal conductivity ratio of $k_m/k_s$=0 (blue), $k_m/k_s$=1/3 (red) and $k_m/k_s$=0 (green), respectively; dashed lines stand for semi-analytical solutions with $R_{ac}$=0. Other parameter values are the same as those in Figure 2.

### 3.2 Effects of thermal conductivity ratio on TER and TOR

We next discuss the effect of the thermal conductivity ratio, $k_m/k_s$, on the TER and TOR under short and open circuit conditions. For typical membrane materials, $k_m/k_s$ is usually of the order of 0.1 to 10 (Table E1). For the sake of generality, a broader $k_m/k_s$ range of $0.01 \leq k_m/k_s \leq 10$ is considered in this study. Evidently, with decreasing $k_m/k_s$, the temperature at the cold end of the nanopore decreases while at the hot end increases (Figure 4a). Accordingly, the effective temperature difference across the nanopore, $\Delta T_p/\Delta T$, increases with decreasing $k_m/k_s$ gradually (Figure 4b). This implies that the decrease in $k_m/k_s$ can give rise to increases in both the SCC and the Seebeck coefficient (Figure 4c to d). Specifically, as $k_m/k_s$ decreases below ~0.02, both the SCC and the Seebeck coefficient tend to plateau values for $k_m/k_s \to 0$. By contrast, as $k_m/k_s$ increases beyond 10, the SCC and the Seebeck coefficient decay to zero since the effective temperature difference, $\Delta T_p$, vanishes.

Despite being shown in the previous section that the adjustable parameter $\beta$ would increase with increasing $k_m/k_s$, we still find that a constant $\beta = 0.036$ could give rise to a fine agreement in the SCC between the numerical and semi-analytical results (Figure 4**c**) in addition to the Seebeck coefficient (Figure 4d). This is because the internal resistance is only weakly dependent on $k_m/k_s$ (Figure 3 and Figure E1), and thus a constant resistance assumption for each $\kappa_\infty a$ could make sense for predicting the SCC based on the semi-analytical model. Moreover, the semi-analytical model can achieve a quite accurate prediction of the TOC under the open circuit condition (Figure 4f), but underestimates the short-circuit TOC magnitude to some degree (Figure 4e**)**. Despite this, our semi-analytical model can provide a first estimation of the short-circuit TOC. Altogether, the semi-analytical model can achieve a quite satisfactory estimation of both the electrical and hydrodynamic parameters.

Significantly, distinctions exist between the TOCs under short- and open-circuit conditions, especially in the signs. Specifically, the TOC under the short-circuit condition is negative and the fluid flow directs to the hot end (Figure 4e). Whereas the TOC under the open-circuit condition displays a positive sign, and thus the flow directs toward the cold end (Figure 4f). This could be understood by noting that the dominant flow components under short circuit conditions — described by the term proportional to the temperature gradient in Eq. (D10) — always direct to the hot end ($L_3<0$; see Eq. (D14)); whereas under open-circuit conditions, the electro-osmotic flow component due to the induced thermoelectric field — whose direction is decided by the sign of $-\Sigma S_e$ — is typically in the opposite direction to the dominant flow components under short circuit conditions but surpass the later in magnitude. Physically, the excess osmotic pressure gradient directly associated with the temperature gradient, $-k_B\sum(n_i-n_v)\partial_x T$, can drive the fluid to the cold end, while the osmotic pressure gradient associated with the ion imbalance inside the EDL [3], $-(e\psi/k_B T)\sum z_i n_i k_B \partial_x T$, and the dielectric force tend to propel the fluid to the hot end. Under the short circuit condition, these driving forces together cause the forward flow in the $+x$ direction. However, under the open circuit condition, the induced thermoelectric field would cause a backflow directing toward the cold end, and such behavior is confirmed by the semi-analytical solution (lines; Figure 4f). This result differs from the previously reported results under the Soret equilibrium [3, 5, 17] in the flow direction since for the latter, the thermo-chemio-osmotic component due to the excess osmotic pressure gradient, $-2k_BT[\cosh(e\psi/k_BT)-1]\partial_x n_v$, could enhance the forward flow [5]. Despite the difference in the signs, the TOC shares the same trend as $\Delta T_p/\Delta T$ whatever the system is under the short/open circuit condition (compare Figure 4c and e; Figure 4d and f). This similarity implies that the same reason accounts for the variation of the TOC with $k_m/k_s$ as that of the SCC or the Seebeck coefficient, that is, the change of $\Delta T_p$.

Finally, we would estimate the magnitude of the TOC. Typically, for either the short or open circuit condition, the TOC, $M_{to}$ falls between $\sim 10^{-10}$ m$^2$ s$^{-1}$ and $\sim 10^{-9}$ m$^2$ s$^{-1}$ or between $O(0.01u_{ref}L)$ and $O(0.1u_{ref}L)$. This range commonly covers the reported experimental results [18] and demonstrates the TOR could be enlarged by reducing the thermal conductivity ratio, $k_m/k_s$.

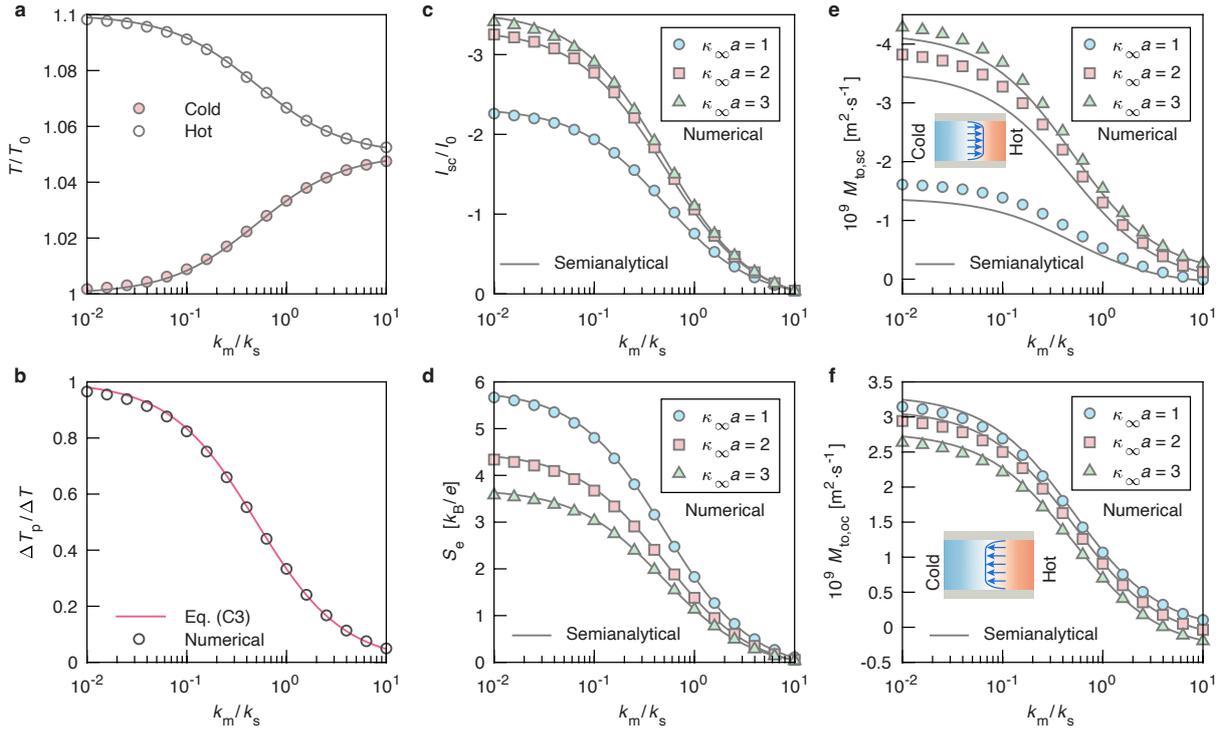

**Figure 4**: **Effects of thermal conductivity ratio on TER and TOR**. (**a**) Dimensionless temperatures at the cold and hot ends of the nanopore as a function of $k_m/k_s$. (**b**) The effective temperature difference across the nanopore, $\Delta T_p$, as a function of the thermal conductivity ratio, $k_m/k_s$. In panels **a** and **b**, Symbols stand for numerical results and lines stand for analytical results (Appendix C). The temperature at the cold or hot end of the nanopore is evaluated by $2\int_0^1 \bar{T}\bar{r}d\bar{r}$. (**c**) Short circuit current (SCC) $I_{sc}$ normalized by $I_0$ as a function of $k_m/k_s$. (**d**) Seebeck coefficient $S_e$ normalized by $k_B/e$ as a function of $k_m/k_s$. (**e**, **f**) Thermoosmotic coefficient $M_{to}$ as a function of $k_m/k_s$ under the short-circuit condition (**e**) and the open circuit condition (**f**). Insets show the flow directions. In panels **c-f**, symbols stand for numerical results, and solid lines stand for results obtained from the semi-analytical model with $\beta =$

0.036 (see Figure E1 for internal resistance). In the calculation, the temperature difference was set to $\Delta \bar{T} = 0.1$, and the surface charge density was set to $\bar{\Sigma}$=50.

*3.3  Effects of surface charge density on TER and TOR*

We then investigate the influences of the surface charge density on the TER and the TOR. To this end, we explore the TER and the TOR at the surface charge densities of $|\bar{\Sigma}|$=10, 30, and 50, which correspond to ~17.86 mC m$^{-2}$, ~53.73 mC m$^{-2}$, and ~89.29 mC m$^{-2}$, respectively. These values fall within the typical range reported in the literature [26, 41, 50]. Our semi-analytical model can provide a quite satisfactory prediction of both the TER and the TOR for various surface charge densities although the prediction of the short-circuit TOC is not quite perfect (Figure 5). Yet, for the short-circuit TOC, our semi-analytical model can still capture the correct trend.

The SCC is found to have a strong dependence on both the magnitude and the sign of the surface charge density. For highly charged nanopores (e.g., $|\bar{\Sigma}|$=30 or 50), whatever the polarity of the surface charge, the SCC, $I_{sc}$, reaches its maximum value at a certain $\kappa_\infty a$ value of ~2 for $|\bar{\Sigma}|$=30 or ~2.75 for $|\bar{\Sigma}|$=50 and decreases once $\kappa_\infty a$ deviates from such a value (Figure 5a). Whereas for low surface charge densities (e.g., $|\bar{\Sigma}|$=10), the SCC displays a more complex behavior. If $\Sigma$>0, $|I_{sc}|$ initially increases and then decreases with increasing $\kappa_\infty a$. In contrast, if $\Sigma$<0, with increasing $\kappa_\infty a$, $I_{sc}$ gradually decreases to zero and switches its sign from positive to negative, then increases in magnitude and eventually saturates to the SCC in the bulk solution (that is, the Soret current caused by the difference in thermodiffusion between cation and anion).

In comparison, the TOC under the short-circuit condition, $M_{to,sc}$, is pronouncedly affected by the magnitude of the surface charge density as well but is less influenced by the polarity of the surface charge, especially for large $\kappa_\infty a$ values (Figure 5b). Clearly, a difference in $M_{to,sc}$ between two surface charge densities of the same magnitude but opposite sign is observed at small $\kappa_\infty a$ values. With the increase of $\kappa_\infty a$, such a difference declines, and eventually the two TOCs converge.

Furthermore, $M_{to,sc}$ arrives at its maximum value at a certain value of $\kappa_\infty a$ and is weakened as $\kappa_\infty a$ deviates from this value. Notably, such a $\kappa_\infty a$ value decreases with decreasing $\bar{\Sigma}$ and is equal to ~3 for $|\bar{\Sigma}|$=50 or ~2 for $|\bar{\Sigma}|$=30 but is beyond our computing range for $|\bar{\Sigma}|$=10.

Next, we focus on the behavior under open-circuit conditions. It is observed from Figure 5c that for positively charged surfaces, the Seebeck coefficient is always positive, and thus the system acts like p-type thermoelectric devices [51]; whereas for negatively charged surfaces, the Seebeck coefficient switches its sign from negative to positive as $\kappa_\infty a$ increases from 1 to 10 (i.e., from n- to p-type [51]). Notably, for a surface charge density of the same magnitude but opposite sign, the Seebeck coefficients for the positive $\bar{\Sigma}$ is usually larger than those for the negative $\bar{\Sigma}$.

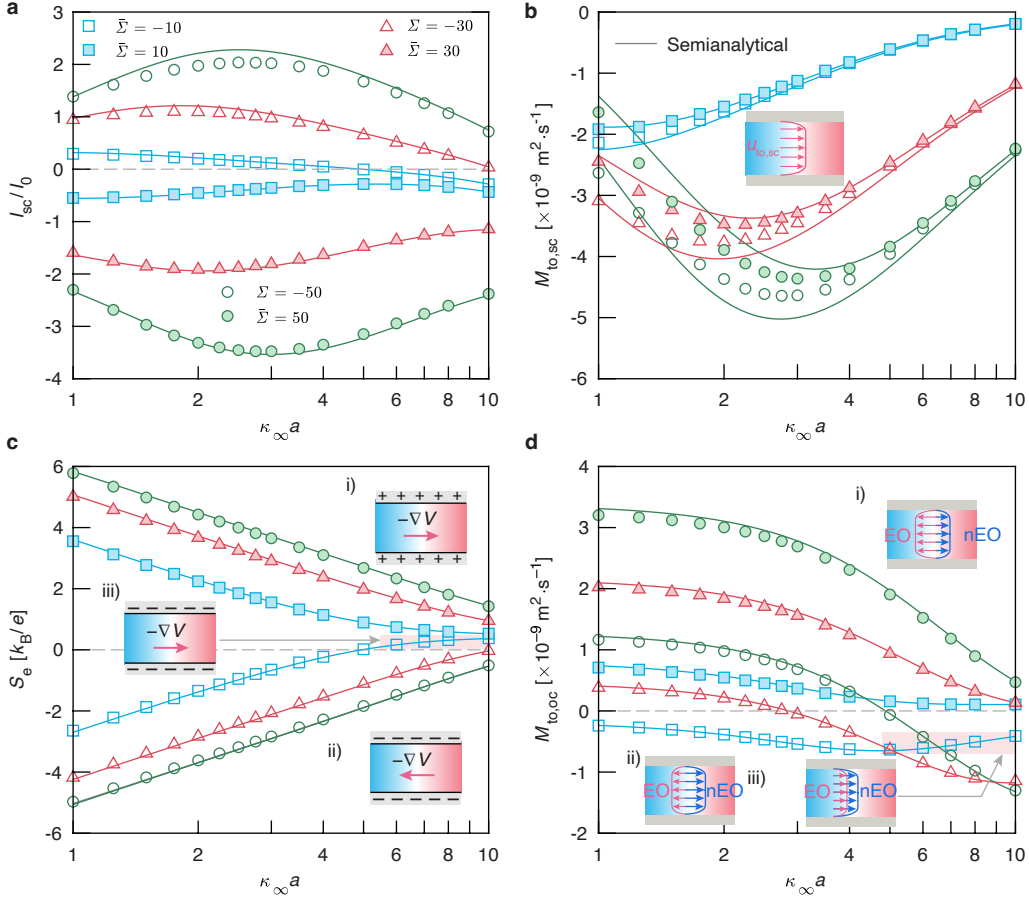

**Figure 5**: **Effects of surface charge density on TER and TOR**. (**a**) Dimensionless short circuit current, $I_{sc}/I_0$, as a function of the Debye parameter, $\kappa_\infty a$, for various surface charge densities, $\bar{\Sigma}$. (**b**) Thermoosmotic coefficient (TOC) $M_{to,sc}$ as a function of the Debye parameter for various surface charge densities under short circuit conditions. The inset shows the flow direction. (**c**) Seebeck coefficient $S_e$ normalized by $k_B/e$ as a function of Debye parameter for various surface charge densities. The insets show the direction of the electric field. (**d**) TOC $M_{to,oc}$ as a function of the Debye parameter for various surface charge densities under open circuit conditions. The insets show the directions of the thermoosmotic flow due to electro-osmosis (EO) and non-electro-osmosis (nEO). The thermal conductivity ratio is set to $k_m/k_s$=0. The thermal conductivity ratio is set to $k_m/k_s$=0. Symbols stand for numerical results; solid lines stand for semi-analytical results (see Fig. E2 for $\beta$ values). In the calculation, the temperature difference was set to $\Delta \bar{T} = 0.1$.

The dependence of the TOC under the open-circuit condition, $M_{to,oc}$, on the surface charge density is more complex (Figure 5d). For positively charged nanopores, $M_{to,oc}$ is always positive at all considered $\kappa_\infty a$ values, indicating that the fluid flows toward the cold end. In addition, the larger the $|\bar{\Sigma}|$ is, the larger the $M_{to,oc}$ becomes. Whereas for negatively charged nanopores, the dependence of $M_{to,oc}$ on $\kappa_\infty a$ is significantly affected by the magnitude of the surface charge density. Specifically, for low $|\bar{\Sigma}|$, $M_{to,oc}$ is always negative and reaches its maximum magnitude at $\kappa_\infty a \approx 5$. For high $|\bar{\Sigma}|$, as $\kappa_\infty a$ increases, $M_{\text{to,oc}}$ decreases gradually to zero and switches sign from positive to negative, then increases reversely in magnitude. This is due to the competition of counteracting factors. First, the TORs excluding the electro-osmotic component caused by the induced electric field share the same direction independently of the polarity of the surface charge (Figure 5b and d, insets). The negative $M_{\text{to,sc}}$ implies that the flow without the electro-osmotic component directs to the hot end (Figure 5b, inset). Conversely, the electro-osmotic component in general directs to the cold end except for large $\kappa_\infty a$ values (blue-ribbon regime; Figure 5d) since the electrostatic body force (whose sign is in accordance with $-\Sigma S_e$ or $\Sigma \Delta V$) points to the cold end (Figure 5c, insets i and ii). The electro-osmotic flow component dominates at small $\kappa_\infty a$ values but becomes weaker and weaker with the increase of $\kappa_\infty a$, and the TOR excluding the electro-osmosis ultimately surpasses the former, leading to the results shown in Fig. 5d.

*3.4 Effects of membrane thickness on TER and TOR*

Finally, we analyze the effects of the membrane thickness (which is reflected by the length-to-radius ratio of the nanopore, $L/a$) on both the TER and the TOR. It is worth mentioning that for relatively thin membranes, the mesh near the nanopore mouths should be refined and for each $L/a$, a mesh independence study has been carried out alone to confirm the appropriate mesh before the calculation. The results for $k_m/k_s=0$ and $k_m/k_s=1/3$ are shown in Figure 6 and Figure 7, respectively. Clearly, the semi-analytical can well reproduce the numerical results for $k_m/k_s=1/3$ (Figure 7) but fails for $k_m/k_s=0$. This is because, for the zero thermal conductivity ratio, Eq. (C3) fails to evaluate the temperature profile at small $L/a$ values. The modification of Eq. (C3) is not the focus of this paper and could be left for the future. Interestingly, with a fitting value for $k_m/k_s = 8 \times 10^{-3}$, the semi-analytical model can well reproduce the numerical results. For $k_m/k_s=0$, as $L/a$ decreases from 100 to 10, the temperature distribution gradually deviates from a linear profile at the nanopore mouth (Figure 6a and Figure C1a). This results in the decrease of the effective temperature difference between two nanopore ends (Figure 6b) and a dramatic increase in the temperature gradient between two nanopore mouths (Figure 6a, inset), which can have a significant effect on the TER and the TOR.

Basically, due to the dramatic decrease in driving force (i.e., the effective temperature gradient, $\Delta T_p/L$; Figure 6a, inset), within the calculation range the SCC is expected to decrease with increasing $L/a$ except for the case of $\kappa_\infty a=1$ (Figure 6c). In such an unusual case, as $L/a$ is increased, the SCC increases first and then decreases. By contrast, with increasing $L/a$, the Seebeck coefficient gradually increases and finally saturates at plateau values regardless of the $\kappa_\infty a$ values (Figure 6d). This is because at small $L/a$, the effective temperature difference across the nanopore, $\Delta T_p$, is smaller than the applied one and gradually increases with increasing $L/a$ (Figure 6b and Figure C1), which tends to enhance the thermally induced potential difference. Consequently, the maximum power density displays a complicated dependence on $L/a$ (Figure 6e). For $\kappa_\infty a = 1$ or 2, the maximum power density reaches its peak value at around $L/a=20$ and gradually decreases when $L/a$ deviates from 20. Whereas for $\kappa_\infty a = 3$ or 5, the maximum power density decreases with $L/a$.

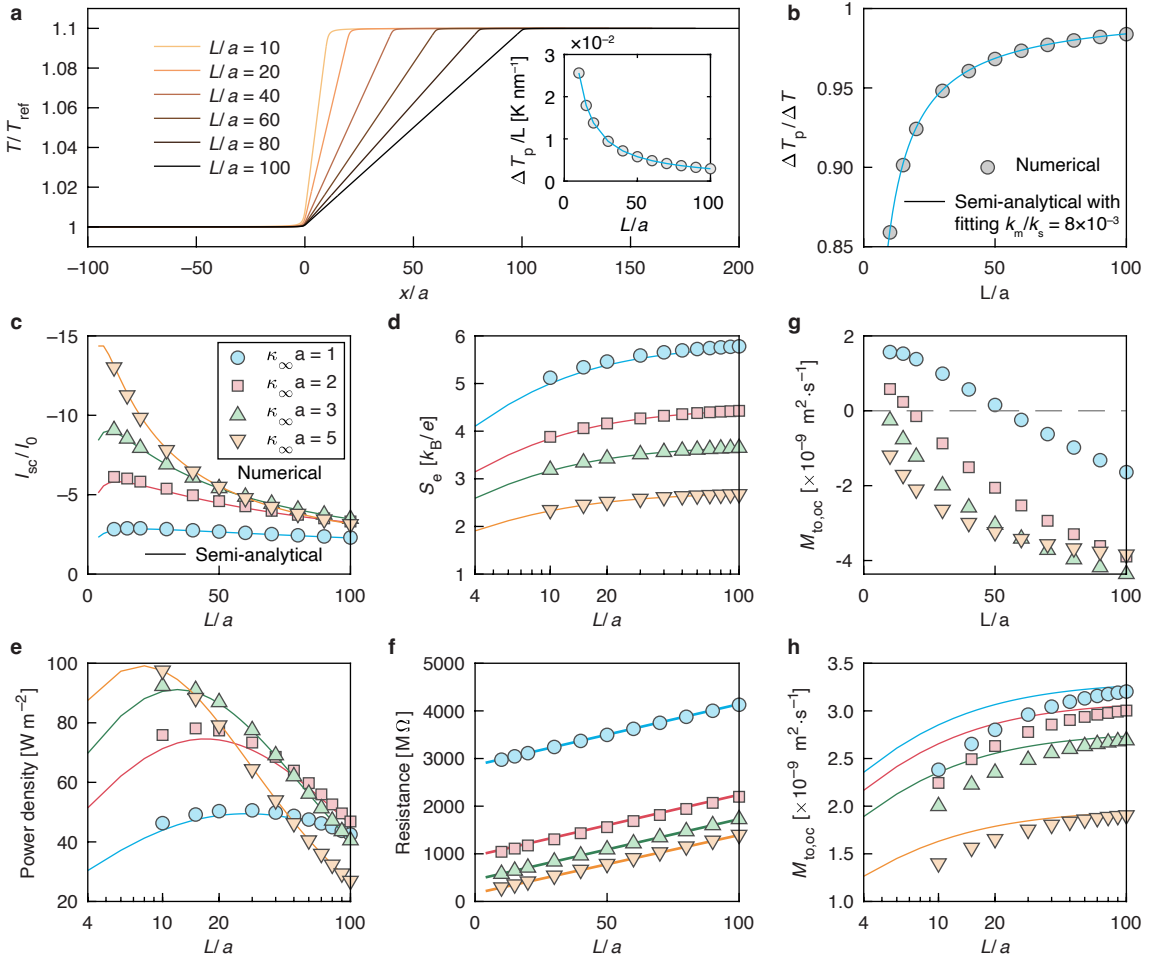

**Figure 6**: **Effects of membrane thickness on TER and TOR at $k_m/k_s=0$**. (**a**) Dimensionless temperature profiles along the central axis for various length-to-radius ratios of the nanopore. Inset shows the dependence of effective temperature gradient on $L/a$. (**b**) The effective temperature difference between two ends of the nanopore, $\Delta T_p$, as a function of $L/a$. (**c-f**) Short circuit current $I_{sc}$ (normalized by $I_0$) (**c**), Seebeck coefficient $S_e$ (normalized by $k_B/e$) (**d**), maximum power density (**e**) and electric resistance (**f**) as functions of length-to-radius ratio, $L/a$, for varying dimensionless Debye parameters. (**g, h**) Thermoosmotic coefficient $M_{to}$ as a function of length-to-radius ratio for varying dimensionless Debye parameters under the short-circuit condition (**g**) and open-circuit condition (**h**). The surface charge density is set to $\overline{\Sigma}=50$, and the thermal conductivity ratio is set to $k_m/k_s=0$. In panels **b** to **h**, symbols stand for numerical results, solid lines are results calculated from the semi-analytical model with a fitting value of $k_m/k_s = 8 \times 10^{-3}$. For evaluating the short circuit current, $\beta = 0.035$.

For all $\kappa_\infty a$ values, the resistance increases linearly with $L/a$, and the corresponding linear fits give non-zero intercepts (Figure 6f). This demonstrates the existence of interface resistance again by noting that the nanopore resistance, $R_p$, is proportional to $L/a$ and the reservoir resistance, $R_{res}$, is not large enough in comparison with the intercepts. Thus, in Figure 6f the intercept stands for $R_{ac} + R_{res}$, and the slope stands for $(2/N_4)/(\pi a \sigma_{ref} Pe)$.

Then, we analyze the dependence of the TOR on the membrane thickness. The TOC under the short circuit condition, $M_{to,sc}$, displays a significant dependence on $L/a$ (Figure 6g). For small $\kappa_\infty a$ values (e.g., $\kappa_\infty a = 1$), with increasing $L/a$, $M_{to,sc}$ decreases gradually from positive to zero, then switches to negative and increases in magnitude. For larger $\kappa_\infty a$ values (such as 5), $M_{to,sc}$ is always negative, and its magnitude increases with increasing $L/a$. In comparison, as $L/a$ increases, the TOC under the open-circuit condition, $M_{to,oc}$, shares the same behavior as the Seebeck coefficient and saturates at a plateau value as $L/a$ is sufficiently large (Figure 6d and Figure 6f). Such a similarity offers a support for the view that for $\kappa_\infty a$ considered here, the TOR due to electro-osmosis is dominated.

For the more practical case of $k_m/k_s=1/3$, the results would be strongly different. Specifically, the linearity of the temperature distribution is better than that for the case of $k_m/k_s=0$ at the nanopore mouths regardless of $L/a$ values (compare Figure 6a, Figure 7a and Figure C1). As $L/a$ decreases from 100 to 10, the effective temperature difference between the two nanopore mouths decreases pronouncedly from $\sim 0.6\Delta T$ to $\sim 0.1\Delta T$ (Figure 7b). Such a decrease is much larger in comparison to the case of $k_m/k_s=0$ (Figure 6b). The variance of the thermal resistance ratio between the reservoirs and the membrane (which is proportional to $L_{res}/L$) with $L/a$ is responsible for this behavior. The change in temperature distribution is expected to cause changes in both TER and TOR. In strong contrast to the case of $k_m/k_s=0$, with increasing $L/a$, the SCC usually increases to the peak value first and then decreases (Figure 7c). The $L/a$ value that the SCC peak arrives at increases with decreasing $\kappa_\infty a$ and is roughly determined as $\sim 40$, $\sim 60$ and $\sim 70$ corresponding to $\kappa_\infty a = 5$, 3 and 2, respectively. Whereas for $\kappa_\infty a = 1$, such a $L/a$ value would be above 100 according to the trend. The variance of the SCC with $L/a$ originates from two counteracting factors. On the one hand, the concentration polarization at two nanopore mouths is mitigated by increasing $L/a$, which tends to increase the SCC. On the other hand, the effective temperature gradient decreases with increasing $L/a$ (Figure 7a, inset), and so does the SCC.

Because of the increase in the effective temperature difference between two nanopore ends with increasing $L/a$, the Seebeck coefficient is expected to increase with increasing $L/a$ as well (Figure 7d). As a result, the maximum power density is expected to display a rather different behavior in contrast to the case of $k_m/k_s$=0 (compare Figure 6e and Figure 7e). Within the $L/a$ range of consideration, the power density increases with increasing $L/a$. In comparison, the resistance increases with $L/a$ linearly regardless of $L/a$ values (Figure 7f), which is identical to the case of $k_m/k_s$=0. However, the intercepts for $k_m/k_s$=1/3 would be larger than their counterparts for $k_m/k_s$=0, implying that the access resistance would increase with $k_m/k_s$, especially for thin membranes.

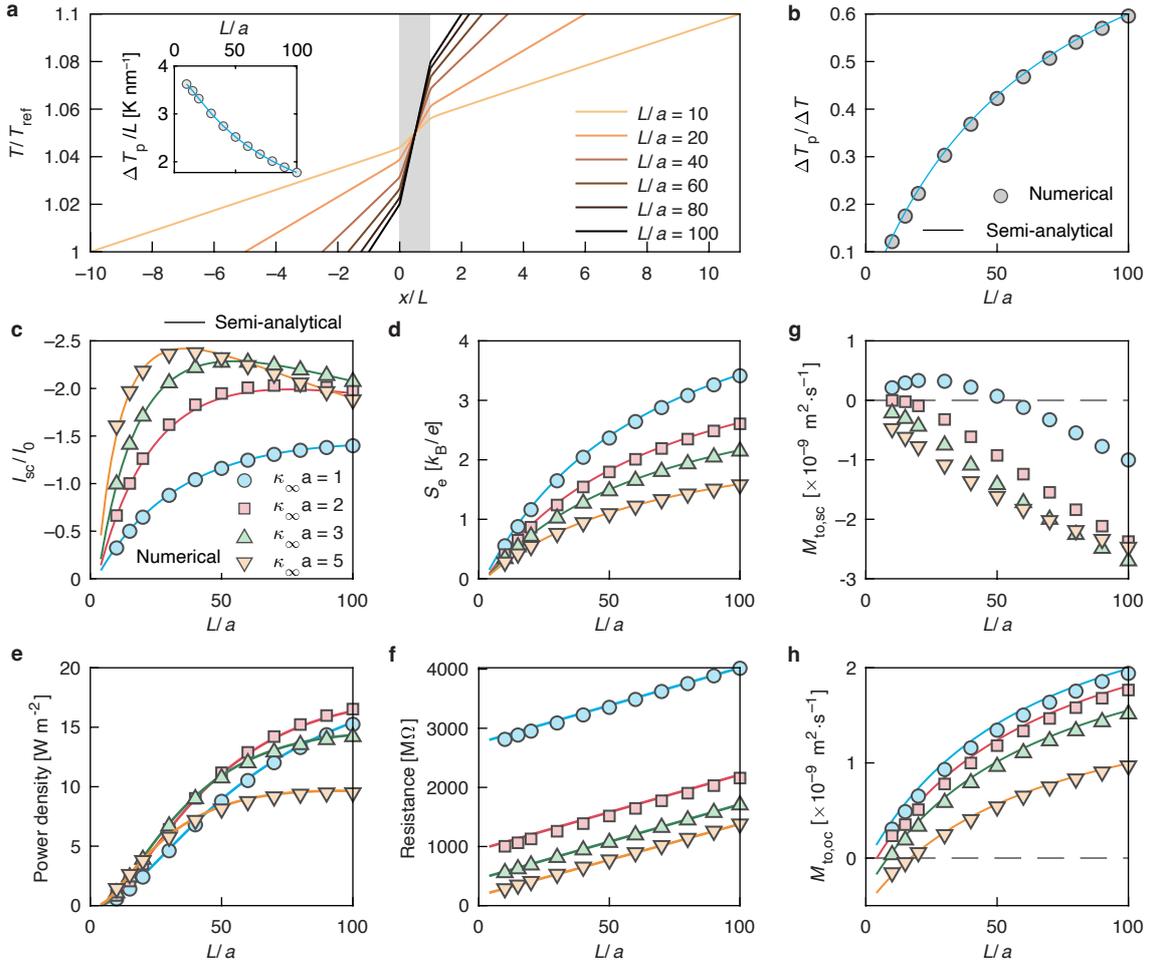

**Figure 7**: **Effects of membrane thickness on TER and TOR at $k_m/k_s$=1/3**. (**a**) Dimensionless temperature profiles along the central axis for various length-to-radius ratios of the nanopore. Inset shows the dependence of effective temperature gradient on $L/a$. (**b**) The effective temperature difference between two ends of the nanopore, $\Delta T_p$, as a function of $L/a$. (**c-f**) Short-circuit current $I_{sc}$ (normalized by $I_0$) (**c**), Seebeck coefficient $S_e$ (in units of $k_B/e$) (**d**), maximum power density (**e**) and resistance (**f**) as functions of length-to-radius ratio, $L/a$, for varying dimensionless Debye parameters. (**g**, **h**) Thermoosmotic coefficient $M_{to}$ as a function of length-to-radius ratio, $L/a$, for varying dimensionless Debye parameters under the short-circuit conditions (**g**) and open-circuit conditions (**h**). All parameter values are the same as Fig. 6 except $k_m/k_s$=1/3. In panels **b** to **h**, symbols stand for numerical results, solid lines are results calculated from the semi-analytical model. For evaluating the short circuit current, $\beta$ = 0.039.

Finally, we discuss the effect of the membrane thickness on the TOR. Overall, the TOC under the short circuit conditions, $M_{to,sc}$, shares an analogous behavior as that in the case of $k_m/k_s=0$ (compare Figure 6g and Figure 7g). Yet, the magnitude of $M_{to,sc}$ is smaller in comparison with the latter. Another difference is that for $\kappa_\infty a =1$, as $L/a$ increases, $M_{to,sc}$ is initially positive and increases first, then decreases and switches sign from positive to negative, then increases in magnitude and eventually is expected to saturate to $u_{to,sc, L/a\to\infty}$ (data not shown). In contrast, the TOC under open-circuit conditions, $M_{to,oc}$, behaves quite differently from that for $k_m/k_s=0$ (compare Figure 6h and Figure 7h), but shows the same trend as the Seebeck coefficient (compare Figure 7d and Figure 7h). A distinct difference is that for $\kappa_\infty a=5$ and $L/a \lesssim 15$, the $M_{to,oc}$ becomes negative. This is because the induced thermoelectric field becomes small enough such that the TOR due to electro-osmosis is surpassed by the other thermoosmotic flow components.

## 4  Conclusions

In this study, we investigated the effects of the fundamental membrane and nanopore parameters (such as the membrane-to-solution thermal conductivity ratio, the surface charge density, and the membrane thickness) on the TER and TOR numerically and semi-analytically. Motivated by the classic space charge model for the osmotic transport of aqueous electrolytes in nanopores [31, 33] or networks of pores [32] under isothermal conditions and the semi-analytical model for the temperature-gradient-induced OCV or/and the open-circuit velocity of nanoconfined electrolyte solutions [3-5], a semi-analytical model is developed for both short circuit parameters (such as the SCC and the short circuit TOC) and open-circuit parameters (such as the Seebeck coefficient and the open circuit TOC) of aqueous electrolyte solutions in membrane nanopores subjected to a temperature gradient. In contrast to the previous models [3-5], the present semi-analytical model not only considers the reservoir/entrance effect but also determines the ion concentration self-consistently using the mass conservation law based on a new calculation strategy. Another advantage of the present model is that it can evaluate the parameters under short-circuit conditions that the previous models cannot. Meanwhile, our semi-analytical model is not only computationally efficient but also can accurately reproduce the predictions of the results obtained from full numerical simulation.

Based on the numerical and semi-analytical results, we found that the SCC arrives at the maximum value when the ratio of the nanopore radius to the Debye length ($\kappa_\infty a$) is ~3 and is prevented from increasing by the access resistance [35, 37] in the EDL heavily overlapped regime. The existence of access resistance can succeed in interpreting experimental results in conical nanochannels quantitatively [41]. Similar behavior is also observed in the streaming current [40]. This, together with the fact that the Seebeck coefficient decreases with the increase of $\kappa_\infty a$, gives rise to an optimum value of $\kappa_\infty a \sim 2$ for maximizing the power density. The maximum power density occurs at the nanopore radius which is twice the Debye length and ranges from several to dozens of W m$^{-2}$, which are translated to several to dozens of mW K$^{-2}$ m$^{-2}$ and are at least an order of magnitude higher than typical thermocells or thermo-supercapacitors [46-49].

In addition, we also showed that a decrease in the thermal conductivity ratio of the membrane can result in a simultaneous enhancement of both TER and TOR by reducing the heat transfer through the supporting membranes. We highlight that under open circuit conditions, the flow direction in our investigated system differs from the analogous systems that are in the Soret equilibrium [3, 5] due to the absence of the thermo-chemio-osmotic flow component caused by the Soret equilibrium. Furthermore, the surface charge sign can heavily affect the sign and magnitude of the SCC, the Seebeck coefficient, and the open circuit TOC but has less effect on the short-circuit TOC. Last, the membrane thickness has made different impacts on the TER and TOR, especially the SCC. An optimum length-to-radius ratio (of serval dozens) is generally observed such that the SCC is maximized for a finite thermal conductivity ratio between the membrane and the aqueous solution, such as 1/3. However, there is no optimum length-to-radius ratio being observed for the open circuit voltage, which is in strong contrast to the results for nanochannels of constant surface potential [6, 23].

In sum, the present study not only develops a new and computationally efficient semi-analytical model for the non-isothermal ion transport in nanoconfined electrolyte solutions but also provides a detailed analysis of the TER and TOR. Our findings are beneficial to an improved understanding of the non-isothermal surface transport in nanofluidic systems and pave a way for their application in the fields of energy conversion and fluid pumping, such as promising thermo-enhanced osmotic energy conversion [52, 53].

Futuristic works on the TER and TOR of nanoconfined electrolyte solutions are expected to be carried out both theoretically and experimentally. For the former, a refined model would be developed to calculate the access resistance self-consistently with higher precision. In addition, other effects, such as the temperature dependence of the surface charge density [54] or zeta potential [55], the slippage on the nanopore wall [36], the activity coefficient correction [56], the surface wettability [57], the ion correlation [58]/finite size effect of ions [59] and the resulting ionic layering behavior [60], can be included in the model development. These effects are of great significance when considering some extreme conditions [5] (such as large temperature differences, highly charged surfaces, high concentration, and so on) or/and special working media (such as ionic liquids or polyelectrolyte solutions). For the experiment aspect, the established model can be modified and improved by a verification experiment conducted based on the well-defined nanopore drilled in a membrane [38]. Furthermore, novel nanofluidic membranes with a small thermal conductivity, a high charge density, and an appropriate ratio of the membrane thickness to the nanopore radius could be synthesized or prepared to enhance TER and TOR.

**Appendix A. Temperature dependencies of physical properties**

The physical properties, such as the dynamic viscosity and the relative permittivity of the electrolyte solution, the ion diffusion coefficients, are dependent on the local temperature in nature. For simplicity, the dependence of the density, the heat capacity at constant pressure, and the thermal conductivity of the electrolyte solution on the temperature are neglected in this work [5]. According to the tabulated data in the literature [61], the dynamic viscosity, the relative permittivity (taken as those of liquid water), and the ion diffusion coefficient are dependent on the temperature:

$$\bar{\eta}(\bar{T}) = \exp\left[-\frac{2.68(\bar{T}-1)+1.1434(\bar{T}-1)^2}{\bar{T}-0.606}\right] \qquad (A1)$$

$$\bar{\varepsilon}_r(\bar{T}) = \exp[-1.3746(\bar{T}-1)] \qquad (A2)$$

$$\bar{D}_i = \bar{T}\bar{\lambda}_i \text{ with } \bar{\lambda}_i = 1 + c_1(\bar{T}-1) + c_2(\bar{T}-1)^2 + c_3(\bar{T}-1)^3 \qquad (A3)$$

where $0.916 \leq \bar{T} \leq 1.251$ (which corresponds to $273.15\ \text{K} \leq T \leq 273.15\ \text{K}$). The coefficients in Eq. (A3) for Na$^+$, K$^+$ and Cl$^-$ are given in **Table A1**.

**Table A1**. Parameter values of Eq. (A3) for simple ions.

| Ion | $c_1$ | $c_2$ | $c_3$ |
|---|---|---|---|
| Na$^+$ | 6.4897 | 8.3576 | −6.0776 |
| Cl$^-$ | 6.0152 | 5.41394 | −4.46065 |
| K$^+$ | 5.7891 | 5.05569 | 0 |

**Appendix B. Streaming potential**

The semi-analytical solution of the streaming potential under isothermal conditions is approximately expressed as [62]

$$(\Delta\bar{\phi})_{\text{str}} = \frac{e(\Delta\phi)_{\text{str}}}{k_B T_{\text{ref}}} = \frac{\Delta\bar{P}}{(\kappa_\infty a)^2} \frac{PeK_1}{PeK_2 + K_3} \qquad (B1)$$

where $Pe=(\varepsilon_0\varepsilon_r/\eta D)[\phi]^2$ is the intrinsic Péclet number [34]; $L_1$, $L_2$ and $L_3$ are respectively expressed as [5]

$$K_1 = \int_0^1 (\bar{\psi}_w - \bar{\psi})\bar{r}\,d\bar{r} \qquad (B2)$$

$$K_2 = \int_0^1 (\bar{\psi}_w - \bar{\psi}) \sinh(\bar{\psi}) \bar{r} d\bar{r} \tag{B3}$$

$$K_3 = \int_0^1 [\cosh(\bar{\psi}) - \chi \sinh(\bar{\psi})] \bar{r} d\bar{r} \tag{B4}$$

with $\bar{\psi} = e\psi/(k_B T_{ref})$ being the dimensionless EDL potential calculated from the well-known Poisson-Boltzmann equation numerically, $\bar{\psi}_w$ being the zeta potential and $\chi$ being the normalized difference in diffusion coefficients between cation and anion denoted by $\chi = (D_+ - D_-)/(D_+ + D_-)$.

**Appendix C. Temperature distribution**

For simplicity, the solution in the nanopore and the membrane are assumed to be in thermal equilibrium (see Figure 1f, top), thus the membrane thermal resistance, $L/k_m\pi(a_{res}^2 - a^2)$, and the nanopore thermal resistance, $L/(k_s\pi a^2)$, are in parallel connection. Then, neglecting the contribution of convective heat transfer, the equivalent thermal resistance of the membrane-nanopore composition can be calculated as

$$\frac{L}{k_{eq}\pi a_{res}^2} = \frac{1}{\frac{k_s\pi a^2}{L} + \frac{k_m\pi(a_{res}^2 - a^2)}{L}} = \frac{L}{\pi k_s a^2} \frac{1}{1 + \frac{k_m}{k_s}\left[\left(\frac{a_{res}}{a}\right)^2 - 1\right]} \tag{C1}$$

The reservoir-membrane/nanopore composition-reservoir is in a series connection (**Fig. 1f**, top), and thus the conservation of the heat flux leads to

$$\frac{\Delta T_p}{\Delta T} = \frac{\frac{L}{k_{eq}}}{\frac{2L_{res}}{k_s} + \frac{L}{k_{eq}}} = \frac{\left(\frac{a_{res}}{a}\right)^2}{\left(\frac{a_{res}}{a}\right)^2 + 2\frac{L_{res}}{L}\left\{1 + \frac{k_m}{k_s}\left[\left(\frac{a_{res}}{a}\right)^2 - 1\right]\right\}} \tag{C2}$$

Assuming the temperature varies linearly in the reservoir or membrane/nanopore, it can be readily found that

$$\bar{T}(\bar{x}) = \begin{cases} 1 + \frac{\Delta\bar{T}}{2}\left(1 - \frac{\Delta T_p}{\Delta T}\right)\frac{\bar{x} + \frac{L_{res}}{a}}{\frac{L_{res}}{a}}, & -\frac{L_{res}}{a} \leq \bar{x} < 0 \\ 1 + \frac{\Delta\bar{T}}{2}\left(1 - \frac{\Delta T_p}{\Delta T}\right) + \Delta\bar{T}\frac{\Delta T_p}{\Delta T}\frac{\bar{x}}{\frac{L}{a}}, & 0 \leq \bar{x} \leq \frac{L}{a} \\ 1 + \frac{\Delta\bar{T}}{2}\left(1 + \frac{\Delta T_p}{\Delta T}\right) + \frac{\Delta\bar{T}}{2}\left(1 - \frac{\Delta T_p}{\Delta T}\right)\frac{\bar{x} - \frac{L}{a}}{\frac{L_{res}}{a}}, & \frac{L}{a} < \bar{x} \leq \frac{L_{res} + L}{a} \end{cases} \tag{C3}$$

It is found from **Figure C1** that at large $L/a$, the temperature distribution in the nanopore could be well described by Eq. (C3). With the increase of $L/a$, the temperature distribution gradually deviates from the Eq. (C3), leading to a decrease in the effective temperature difference across the nanopore. In addition, at small $L/a$, Eq. (C3) could capture a better consistency between numerical and analytical results for $k_m/k_s=1/3$ as compared to $k_m/k_s=0$.

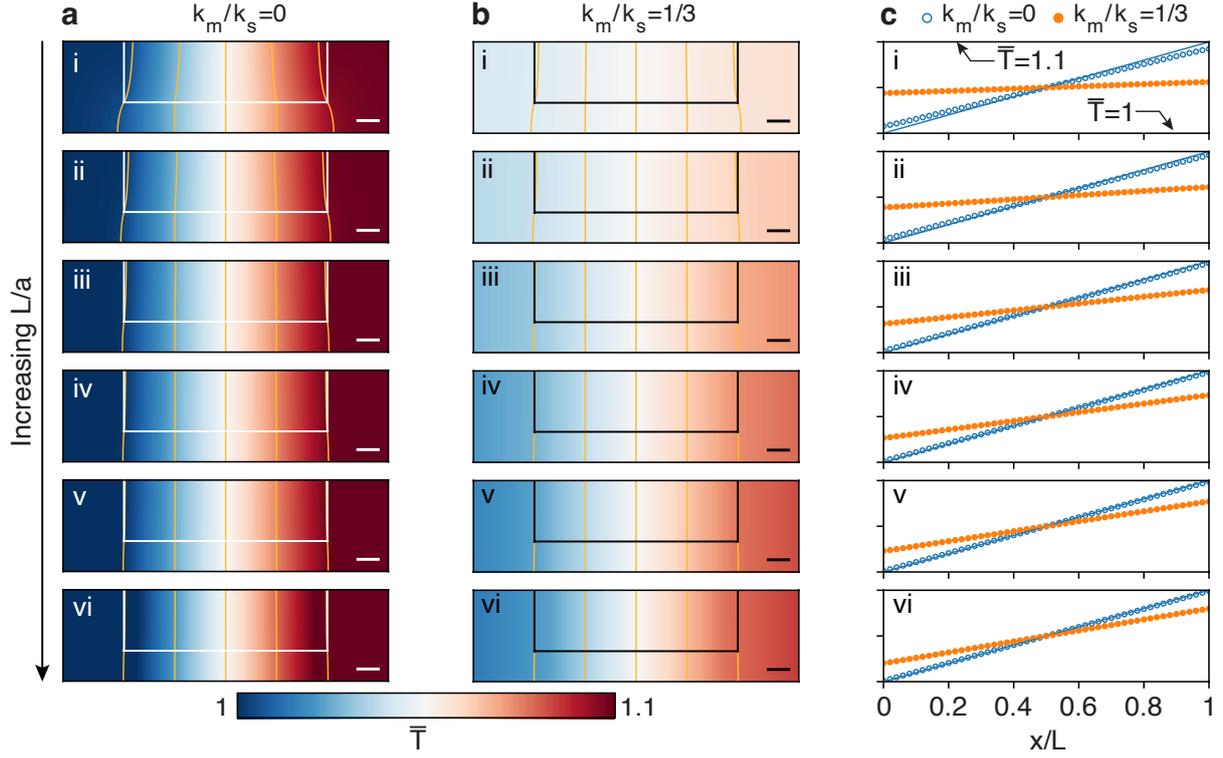

**Figure C1**: Dimensionless temperature distribution at (**a**) $k_m/k_s$=0 and (**b**) $k_m/k_s$ =1/3 for a nanopore length-to-radius ratio of (i) $L/a$=10, (ii) $L/a$=20, (iii) $L/a$=40, (iv) $L/a$=60, (v) $L/a$=80 and (vi) $L/a$=100. Scale bars stand for $L/10$ in axial direction and colored lines stand for isotherms. All results are obtained at $\bar{\Sigma}$=50 and $\kappa_\infty a$=1 using numerical simulation. In addition, the short-circuit condition is considered in the calculations. (**c**) Dimensionless temperature profile along the central axis of the nanopore. Lines are obtained from Eq. (C3), and symbols stand for the numerical results.

**Appendix D. Derivation and numerical approach of the semi-analytical model[3]**

The net electric current in the nanopore can be calculated by

$$I = 2\pi e \int_0^a r(J_{+,x} - J_{-,x})\mathrm{d}r \tag{D1}$$

where $J_{\pm,x}$ is the $x$ component of the ionic flux denoted as

$$J_{\pm,x} = n_v \exp\left(\mp\frac{e\psi}{k_B T}\right)\left\{u_x - D_\pm\left[\frac{\partial n_v}{\partial x} + \left(\frac{2\alpha_\pm}{T} \pm \frac{e\psi}{k_B T^2}\right)\frac{\partial T}{\partial x} \pm \frac{e}{k_B T}\frac{\partial \phi_v}{\partial x}\right]\right\} \tag{D2}$$

in which the celebrated Boltzmann distribution for ions, $n_\pm = n_v\exp(\mp e\psi/k_B T)$, is employed [5], The assumption of equal pre-factor $n_v$ for either ion species has been shown to provide an excellent approximation for estimating observable quantities for the system consisting of a nanochannel and two relatively large reservoirs at both ends [63].

Under the condition of $a/L \ll 1$, following our previous work [5] and making use of Eqs. (D1) and (D2), the electric current is derived as

---

[3] The PDF version downloaded from publisher's website contains some typos in symbols used in this Appendix although the online version does not contains these typos. In this authors' manuscript, the symbols are the same as the publisher's online version.

$$\bar{I} = N_1\left(-\frac{\partial \bar{p}_v}{\partial \bar{x}}\right) + N_2\left(-\frac{\partial \bar{n}_v}{\partial \bar{x}}\right) + N_3\left(-\frac{\partial \bar{T}}{\partial \bar{x}}\right) + N_4\left(-\frac{\partial \bar{\phi}_v}{\partial \bar{x}}\right) \tag{D3}$$

where

$$N_1 = \frac{4}{\bar{\eta}} \frac{\bar{\varepsilon}_r \bar{T}}{(\kappa_\infty a)^2} \int_0^1 (\Psi - \Psi_w) \bar{r}\, d\bar{r} \tag{D4}$$

$$N_2 = \frac{4\bar{\varepsilon}_r \bar{T}^2}{\bar{\eta}} \int_0^1 (\Psi - \Psi_w)[\cosh(\Psi) - 1]\bar{r}\, d\bar{r} - \frac{4\bar{D}}{Pe} \int_0^1 [\sinh(\Psi) - \chi \cosh(\Psi)]\bar{r}\, d\bar{r} \tag{D5}$$

$$N_3 = \frac{4\bar{\varepsilon}_r \bar{n}_v \bar{T}}{\bar{\eta}} \int_0^1 (\Psi - \Psi_w)\left[\cosh(\Psi) - 1 - \Psi \sinh(\Psi) + \frac{\gamma^*}{2} \frac{\bar{\varepsilon}_r \bar{T}^2}{\bar{n}_v}\left(\frac{1}{\kappa_\infty a}\frac{\partial \Psi}{\partial \bar{r}}\right)^2\right]\bar{r}\, d\bar{r} +$$

$$\frac{4\bar{D}}{Pe}\frac{\bar{n}_v}{\bar{T}}\left\{-2\alpha \int_0^1 [\sinh(\Psi) - \gamma \cosh(\Psi)]\bar{r}\, d\bar{r} + \int_0^1 \Psi[\cosh(\Psi) - \chi \sinh(\Psi)]\bar{r}\, d\bar{r}\right\} \tag{D6}$$

$$N_4 = -\frac{4\bar{\varepsilon}_r \bar{n}_v \bar{T}}{\bar{\eta}} \int_0^1 (\Psi - \Psi_w)\sinh(\Psi)\bar{r}\, d\bar{r} + \frac{4\bar{D}}{Pe}\frac{\bar{n}_v}{\bar{T}} \int_0^1 [\cosh(\Psi) - \chi \sinh(\Psi)]\bar{r}\, d\bar{r} \tag{D7}$$

with $\gamma^* = T_{\text{ref}}(\partial_T \varepsilon_r / \varepsilon_r)$ being the reduced temperature sensibility of the solution relative permittivity, $\bar{D} = \sum_i D_i / \sum_i D_{i,\text{ref}}$ the average ion diffusivity, $\alpha = \sum_i \alpha_i D_i / \sum_i D_i$ the reduced Soret coefficient of ions, $\gamma = (\alpha_+ D_+ - \alpha_- D_-)/\sum_i \alpha_i D_i$ the normalized difference in the diffusivity-modified Soret coefficients between cation and anion, and $\Psi = e\psi/(k_B T)$ the local EDL potential calculated by

$$\frac{1}{\bar{r}}\frac{\partial}{\partial \bar{r}}\left(\bar{r}\frac{\partial \Psi}{\partial \bar{r}}\right) = \frac{(\kappa_\infty a)^2 \bar{n}_v}{\bar{\varepsilon}_r \bar{T}}\sinh(\Psi) \tag{D8}$$

$$\left.\frac{\partial \Psi}{\partial \bar{r}}\right|_{\bar{r}=0} = 0,\quad \left.\frac{\partial \Psi}{\partial \bar{r}}\right|_{\bar{r}=1} = \frac{\bar{\Sigma}}{\bar{\varepsilon}_r \bar{T}} \tag{D9}$$

For closing Eq. (D3), the average velocity and the average ion flux are derived respectively as

$$\bar{u}_{\text{to}} = -(\partial_{\bar{x}} \bar{p}_v) L_1 - (\partial_{\bar{x}} \bar{n}_v) L_2 - (\partial_{\bar{x}} \bar{T}) L_3 - (\partial_{\bar{x}} \bar{\phi}_v) L_4 \tag{D10}$$

$$\bar{J}_{\text{ion}} = -(\partial_{\bar{x}} \bar{p}_v) M_1 - (\partial_{\bar{x}} \bar{n}_v) M_2 - (\partial_{\bar{x}} \bar{T}) M_3 - (\partial_{\bar{x}} \bar{\phi}_v) M_4 \tag{D11}$$

where

$$L_1 = \frac{1}{8\bar{\eta}} \tag{D12}$$

$$L_2 = \frac{(\kappa_\infty a)^2}{2}\frac{\bar{T}}{\bar{\eta}} \int_0^1 [\cosh(\Psi) - 1](1 - \bar{r}^2)\bar{r}\, d\bar{r} \tag{D13}$$

$$L_3 = \frac{(\kappa_\infty a)^2}{2}\frac{\bar{n}_v}{\bar{\eta}} \int_0^1 \left[\cosh(\Psi) - 1 - \Psi \sinh(\Psi) + \frac{\gamma^*}{2}\frac{\bar{\varepsilon}_r \bar{T}^2}{\bar{n}_v}\left(\frac{1}{\kappa_\infty a}\frac{\partial \Psi}{\partial \bar{r}}\right)^2\right](1 - \bar{r}^2)\bar{r}\, d\bar{r} \tag{D14}$$

$$L_4 = \frac{2\bar{\varepsilon}_r \bar{T}}{\bar{\eta}} \int_0^1 (\Psi - \Psi_w)\bar{r}\, d\bar{r} \tag{D15}$$

$$M_1 = \frac{\bar{n}_v}{\bar{\eta}} \int_0^1 \cosh(\Psi)(1 - \bar{r}^2)\bar{r}\, d\bar{r} \tag{D16}$$

$$M_2 = -\frac{(\kappa_\infty a)^2}{2}\frac{4\bar{n}_v \bar{T}}{\bar{\eta}} \int_0^1 \left\{\cosh(\Psi)\left[\cosh(\Psi) - 1 - \frac{\bar{\varepsilon}_r \bar{T}}{2\bar{n}_v}\left(\frac{1}{\kappa_\infty a}\frac{\partial \Psi}{\partial \bar{r}}\right)^2\right] + [\cosh(\Psi) -$$

$$1]\left[\cosh(\Psi) - \frac{\bar{\varepsilon}_r \bar{T}}{2\bar{n}_v}\left(\frac{1}{\kappa_\infty a}\frac{\partial \Psi}{\partial \bar{r}}\right)^2\right]\right\}\bar{r}^3 \ln \bar{r}\, d\bar{r} + \frac{4\bar{D}}{Pe} \int_0^1 [\cosh(\Psi) - \chi \sinh(\Psi)]\bar{r}\, d\bar{r} \tag{D17}$$

$$M_3 = -\frac{(\kappa_\infty a)^2}{2} \frac{4\bar{n}_v^2}{\bar{\eta}} \Big\{ 2\int_0^1 \cosh(\Psi)\,\bar{r}\ln\bar{r}\,d\bar{r} \int_0^{\bar{r}} \Big[\cosh(\Psi) - 1 - \Psi\sinh(\Psi) +$$

$$\frac{\gamma^* \bar{\varepsilon}_r \bar{T}^2}{2} \frac{1}{\bar{n}_v}\left(\frac{1}{\kappa_\infty a}\frac{\partial\Psi}{\partial\hat{r}}\right)^2 \Big]\hat{r}d\hat{r} + \int_0^1 \Big[\cosh(\Psi) - 1 - \Psi\sinh(\Psi) + \frac{\gamma^* \bar{\varepsilon}_r \bar{T}^2}{2}\frac{1}{\bar{n}_v}\left(\frac{1}{\kappa_\infty a}\frac{\partial\Psi}{\partial\bar{r}}\right)^2\Big]\Big[\cosh(\Psi) -$$

$$\frac{\bar{\varepsilon}_r \bar{T}}{2\bar{n}_v}\left(\frac{\partial_{\bar{r}}\Psi}{\kappa_\infty a}\right)^2\Big]\bar{r}^3\ln\bar{r}\,d\bar{r}\Big\} + \frac{4\bar{D}}{Pe}\frac{\bar{n}_v}{\bar{T}}\Big\{2\alpha\int_0^1[\cosh(\Psi) - \gamma\sinh(\Psi)]\bar{r}d\bar{r} - \int_0^1 \Psi[\sinh(\Psi) -$$

$$\chi\cosh(\Psi)]\bar{r}d\bar{r}\Big\} \tag{D18}$$

$$M_4 = \frac{4\bar{\varepsilon}_r\bar{n}_v\bar{T}}{\bar{\eta}}\int_0^1 (\Psi-\Psi_w)\cosh(\Psi)\bar{r}d\bar{r} - \frac{4\bar{D}}{Pe}\frac{\bar{n}_v}{\bar{T}}\int_0^1[\sinh(\Psi) - \chi\cosh(\Psi)]\bar{r}d\bar{r} \tag{D19}$$

Further, the flow rate, the current and the overall ion flux in the reservoirs are deduced from Eqs. (D10), (D3) and (D11), respectively, as

$$\bar{u}_{to} = -\frac{a_{res}^2}{a^2}\frac{1}{8\bar{\eta}}\frac{\partial\bar{p}_v}{\partial\bar{x}} \tag{D20}$$

$$\bar{J}_{ion} = -\frac{a_{res}^2}{a^2}\frac{\bar{n}_v}{4\bar{\eta}}\frac{\partial\bar{p}_v}{\partial\bar{x}} - \frac{a_{res}^2}{a^2}\frac{2\bar{D}}{Pe}\left[\frac{\partial\bar{n}_v}{\partial\bar{x}} + 2\alpha\frac{\bar{n}_v}{\bar{T}}\frac{\partial\bar{T}}{\partial\bar{x}} + \chi\frac{\bar{n}_v}{\bar{T}}\frac{\partial\bar{\phi}_v}{\partial\bar{x}}\right] \tag{D21}$$

$$\bar{I} = -\frac{a_{res}^2}{a^2}\frac{2\bar{D}}{Pe}\left[\chi\frac{\partial\bar{n}_v}{\partial\bar{x}} + 2\gamma\alpha\frac{\bar{n}_v}{\bar{T}}\frac{\partial\bar{T}}{\partial\bar{x}} + \frac{\bar{n}_v}{\bar{T}}\frac{\partial\bar{\phi}_v}{\partial\bar{x}}\right] \tag{D22}$$

It follows from Eqs. (D3) and (D22) that

$$\frac{\partial\bar{\phi}_v}{\partial\bar{x}} = \begin{cases} -\frac{a^2}{a_{res}^2}\frac{\bar{T}}{\bar{n}_v}\frac{Pe}{2\bar{D}}\bar{I} - \chi\frac{\bar{T}}{\bar{n}_v}\frac{\partial\bar{n}_v}{\partial\bar{x}} - 2\gamma\alpha\frac{\partial\bar{T}}{\partial\bar{x}}, & -\frac{L_{res}}{a} \le \bar{x} < 0 \text{ or } \frac{L}{a} < \bar{x} \le \frac{L+L_{res}}{a} \\ -\frac{\bar{I}}{N_4} - \frac{N_1}{N_4}\frac{\partial\bar{P}_v}{\partial\bar{x}} - \frac{N_2}{N_4}\frac{\partial\bar{n}_v}{\partial\bar{x}} - \frac{N_3}{N_4}\frac{\partial\bar{T}}{\partial\bar{x}}, & 0 \le \bar{x} \le \frac{L}{a} \end{cases} \tag{D23}$$

Substituting Eq. (D23) into Eqs. (D10), (D11), (D21) and (D22) yields

$$\frac{\partial\bar{p}_v}{\partial\bar{x}} = \begin{cases} -8\bar{\eta}\bar{u}_{to}\frac{a^2}{a_{res}^2}, & -\frac{L_{res}}{a} \le \bar{x} < 0 \text{ or } \frac{L}{a} < \bar{x} \le \frac{L+L_{res}}{a} \\ \frac{N_4\bar{u}_{to}}{L_4N_1-L_1N_4} - \frac{L_4\bar{I}}{L_4N_1-L_1N_4} - \frac{L_4N_2-L_2N_4}{L_4N_1-L_1N_4}\frac{\partial\bar{n}_v}{\partial\bar{x}} - \frac{L_4N_3-L_3N_4}{L_4N_1-L_1N_4}\frac{\partial\bar{T}}{\partial\bar{x}}, & 0 \le \bar{x} \le \frac{L}{a} \end{cases} \tag{D24}$$

$$\frac{\partial\bar{n}_v}{\partial\bar{x}} =$$

$$\begin{cases} -\frac{\bar{J}_{ion}}{1-\chi^2}\frac{a^2}{a_{res}^2}\frac{Pe}{2\bar{D}} + \frac{\chi\bar{I}}{1-\chi^2}\frac{a^2}{a_{res}^2}\frac{Pe}{2\bar{D}} - \frac{1}{1-\chi^2}\frac{\bar{n}_v}{4\bar{\eta}}\frac{Pe}{2\bar{D}}\frac{\partial\bar{p}_v}{\partial\bar{x}} - 2\alpha\frac{1-\chi\gamma}{1-\chi^2}\frac{\bar{n}_v}{\bar{T}}\frac{\partial\bar{T}}{\partial\bar{x}}, & -\frac{L_{res}}{a} \le \bar{x} < 0 \text{ or } \frac{L}{a} < \bar{x} \le \frac{L+L_{res}}{a} \\ \frac{N_4\bar{J}_{ion}}{M_4N_2-M_2N_4} - \frac{M_4\bar{I}}{M_4N_2-M_2N_4} - \frac{M_4N_1-M_1N_4}{M_4N_2-M_2N_4}\frac{\partial\bar{P}_v}{\partial\bar{x}} - \frac{M_4N_3-M_3N_4}{M_4N_2-M_2N_4}\frac{\partial\bar{T}}{\partial\bar{x}}, & 0 \le \bar{x} \le \frac{L}{a} \end{cases}$$

$$\tag{D25}$$

Integrating Eqs. (D24) to (D25) from cold to hot end and noting that fluxes ($\bar{I}, \bar{u}_{to}, \bar{J}_{ion}$) are invariant in $x$ direction and $\Delta\bar{P}_v = 0$, $\Delta\bar{n}_v=0$, one obtains

$$\bar{u}_{to} = \left[-8\frac{a^2}{a_{res}^2}\left(\int_{-\frac{L_{res}}{a}}^{0}\bar{\eta}d\bar{x} + \int_{\frac{L}{a}}^{\frac{L+L_{res}}{a}}\bar{\eta}d\bar{x}\right) + \int_0^{\frac{L}{a}}\frac{N_4 d\bar{x}}{L_4N_1-L_1N_4}\right]^{-1} \times \Big[\bar{I}\int_0^{\frac{L}{a}}\frac{L_4 d\bar{x}}{L_4N_1-L_1N_4} +$$

$$\int_0^{\frac{L}{a}}\frac{L_4N_2-L_2N_4}{L_4N_1-L_1N_4}\frac{\partial\bar{n}_v}{\partial\bar{x}}d\bar{x} + \int_0^{\frac{L}{a}}\frac{L_4N_3-L_3N_4}{L_4N_1-L_1N_4}\frac{\partial\bar{T}}{\partial\bar{x}}d\bar{x}\Big] \tag{D26}$$

$$\bar{J}_{\text{ion}} = \left[ -\frac{a^2}{a_{\text{res}}^2} \left( \int_{-\frac{L_{\text{res}}}{a}}^{0} \frac{Pe}{2\bar{D}} \frac{d\bar{x}}{1-\chi^2} + \int_{\frac{L}{a}}^{\frac{L}{a}+\frac{L_{\text{res}}}{a}} \frac{Pe}{2\bar{D}} \frac{d\bar{x}}{1-\chi^2} \right) + \int_{0}^{\frac{L}{a}} \frac{N_4 d\bar{x}}{M_4 N_2 - M_2 N_4} \right]^{-1} \times$$

$$\left\{ \bar{I} \left[ -\frac{a^2}{a_{\text{res}}^2} \left( \int_{-\frac{L_{\text{res}}}{a}}^{0} \frac{Pe}{2\bar{D}} \frac{\chi}{1-\chi^2} d\bar{x} + \int_{\frac{L}{a}}^{\frac{L}{a}+\frac{L_{\text{res}}}{a}} \frac{Pe}{2\bar{D}} \frac{\chi}{1-\chi^2} d\bar{x} \right) + \int_{0}^{\frac{L}{a}} \frac{M_4 d\bar{x}}{M_4 N_2 - M_2 N_4} \right] + \int_{-\frac{L_{\text{res}}}{a}}^{0} \frac{\bar{n}_v}{4\bar{\eta}} \frac{Pe}{2\bar{D}} \frac{\partial \bar{p}_v}{\partial \bar{x}} \frac{d\bar{x}}{1-\chi^2} +$$

$$2 \int_{-\frac{L_{\text{res}}}{a}}^{0} \alpha \frac{1-\chi\gamma}{1-\chi^2} \frac{\bar{n}_v}{\bar{T}} \frac{\partial \bar{T}}{\partial \bar{x}} d\bar{x} + \int_{0}^{\frac{L}{a}} \frac{M_4 N_1 - M_1 N_4}{M_4 N_2 - M_2 N_4} \frac{\partial \bar{p}_v}{\partial \bar{x}} d\bar{x} + \int_{0}^{\frac{L}{a}} \frac{M_4 N_3 - M_3 N_4}{M_4 N_2 - M_2 N_4} \frac{\partial \bar{T}}{\partial \bar{x}} d\bar{x} +$$

$$\left. \int_{\frac{L}{a}}^{\frac{L}{a}+\frac{L_{\text{res}}}{a}} \frac{\bar{n}_v}{4\bar{\eta}} \frac{Pe}{2\bar{D}} \frac{\partial \bar{p}_v}{\partial \bar{x}} \frac{d\bar{x}}{1-\chi^2} + 2 \int_{\frac{L}{a}}^{\frac{L}{a}+\frac{L_{\text{res}}}{a}} \alpha \frac{1-\chi\gamma}{1-\chi^2} \frac{\bar{n}_v}{\bar{T}} \frac{\partial \bar{T}}{\partial \bar{x}} d\bar{x} \right\} \quad \text{(D27)}$$

Under the open circuit condition, the current, $\bar{I}$, vanishes, and thus the developed semi-analytical model can be performed by Algorithm 1. In this study, this algorithm is carried out using COMSOL LiveLink$^{\text{TM}}$ for MATLAB. Specifically, Eqs. (D8) and (D9) are solved numerically by the general PDE module, and integrals $L_j$, $M_j$, and $N_j$ are evaluated by the built-in algorithm in COMSOL. Furthermore, the iteration is implemented by MATLAB routines.

To obtain the SCC and the short-circuit TOC, we analyze the internal resistance of the investigated system as follows. As shown in the bottom panel of **Figure 1f**, the total resistance is comprised of three resistances in series, that is, the reservoir resistance, the nanopore resistance and the access resistance. Among them, the reservoir resistance is naturally given as

$$R_{\text{res}} = \int_{-L_{\text{res}}}^{0} \frac{k_B T}{e^2 n_\infty (D_+ + D_-)} \frac{dx}{\pi a_{\text{res}}^2} + \int_{L+L_{\text{res}}}^{L+2L_{\text{res}}} \frac{k_B T}{e^2 n_\infty (D_+ + D_-)} \frac{dx}{\pi a_r^2} \quad \text{(D28)}$$

which can be rewritten as Eq. (19) in the main text. Clearly, the enhancement of the nanopore conductance due to the surface charge and the electroosmosis is reflected by the factor $N_4$, and thus the nanopore resistance can be readily derived as Eq. (20). Following Ref. [37], the access resistance is given as

$$R_{\text{ac}} = \frac{1}{\sigma_{\text{ref}}} \left[ \frac{1}{4a + \beta l_{\text{Du},\bar{x}=0}} + \frac{1}{4a + \beta l_{\text{Du},\bar{x}=L/a}} \right] \quad \text{(D29)}$$

which discards the assumption of equal Dukhin length ($l_{\text{Du}}$) at two nanopore mouths because of the difference in local temperature. Neglecting the advective effect and equal ion diffusivity assumption, the Dukhin length was derived as [37]

$$l_{\text{Du}} = \frac{a}{2} \left[ \frac{1}{\pi a^2 n_\infty (D_+ + D_-)} \int_0^a (D_+ n_+ + D_- n_-) r dr - 1 \right] \quad \text{(D30)}$$

Substituting $n_\pm = n_v \exp(\mp e\psi/k_B T)$ into Eq. (D30), we can readily obtain Eq. (22).

The numerical approach for calculating the SCC and the short-circuit TOC is briefly described as follows. First, $\beta$ is determined by minimizing the sum of squared error between the total resistance (or the SCC) obtained from numerical simulation and Eq. (23) (or Eq. (18)). For the latter, the results obtained from Algorithm 1, such as $n_v$, are used to calculate the resistance.

Subsequently, the SCC is evaluated by Eq. (18) with the OCV obtained by Algorithm 1. Furthermore, the short-circuit TOC is evaluated by Eq. (D26).

| | |
|---|---|
| **Algorithm 1**: Numerical algorithm of the semi-analytical model under the open-circuit condition. For simplicity, the bars over the notations are omitted here. In this study, the relative tolerance is set to $\epsilon=10^{-5}$, $N$ is set to 151. | |
| **Step 1** | Calculate $T_i=T(x_i)$ by Eq. (C3) at discrete points $-L_{res}/a=x_0 \leq x_1 \leq ... \leq x_N = L/a + L_{res}/a$ |
| **Step 2** | Initialize $n_{v,i}=n(x_i)=1$ and $\varphi_i=\varphi(T_i)$, where $\varphi=\eta, \varepsilon_r, D, \chi, \alpha, \gamma$, and $\gamma^*$ |
| **Step 3** | Solve Eqs. (D8) and (D9) numerically and calculate integrals $L_j(x_i)$, $M_j(x_i)$ and $N_j(x_i)$ |
| **Step 4** | Calculate $u_{to}$ by Eq. (D26) and then calculate $p_v$ by integration of Eq. (D24) with $p_v(-L_{res}/a)=0$ |
| **Step 5** | Calculate $J_{ion}$ by Eq. (D27) and then calculate $n_v$ by the integration of Eq. (D25) with $n_v(-L_{res}/a)=1$ |
| **Step 6** | Calculate $\varphi_v$ by integration of Eq. (D23) with $\varphi_v(-L_{res}/a)=0$ |
| **Step 7** | Iterate Steps 3-6 until the following tolerances are satisfied: $\left|(u_{to}-u_{to}^{old})/u_{to}^{old}\right|<\epsilon$, $\left|(J_{ion}-J_{ion}^{old})/J_{ion}^{old}\right|<\epsilon$ and $\max_i \left|(n_{v,i}-n_{v,i}^{old})/n_{v,i}^{old}\right|<\epsilon$ |

## Appendix E. Extended data figures and tables

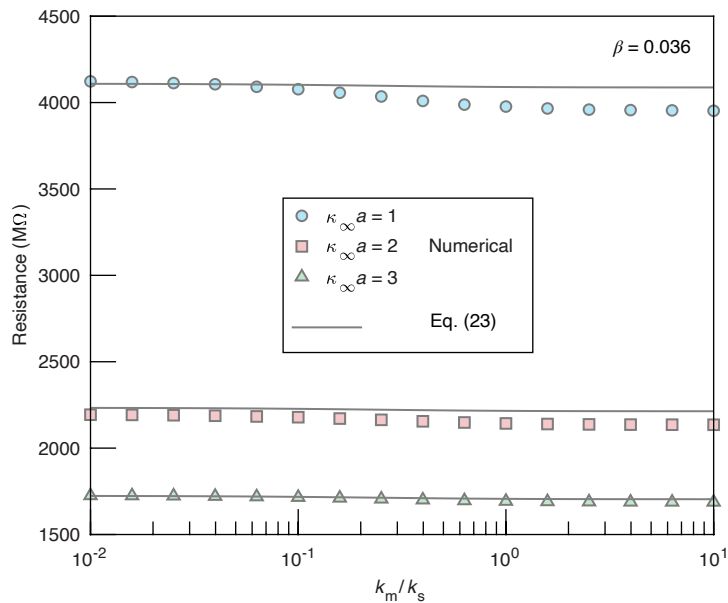

**Figure E1**: Total internal resistance as a function of thermal conductivity ratio. The parameter values are the same as those in Fig. 4.

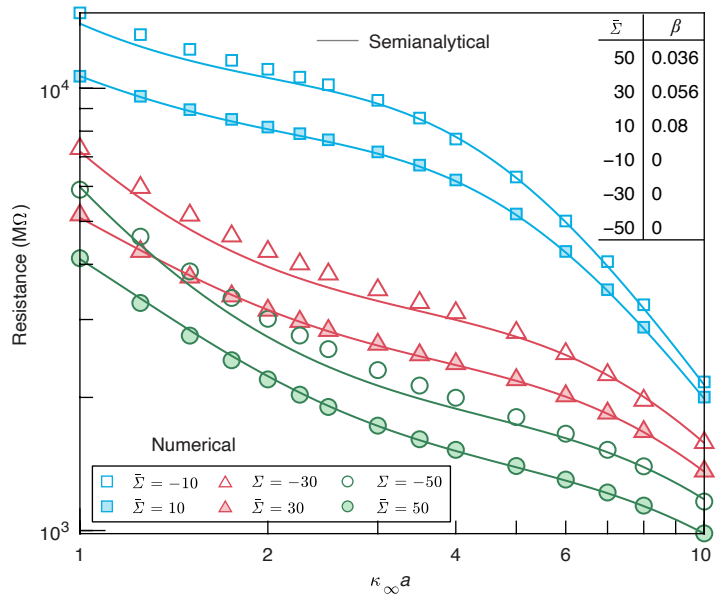

**Figure E2**: Total internal resistance as a function of the Debye parameter for various surface charge densities $\bar{\Sigma}$ at $k_m/k_s=0$. Values of the adjustable parameter $\beta$ for various $\bar{\Sigma}$ is sown in the plot. The parameter values are the same as those in Fig. 5.

**Table E1**: Thermal conductivity of typical membrane materials and the corresponding thermal conductivity ratio.

| Membrane materials | Temperature (°C) | $k_m$ (W m$^{-1}$ K$^{-1}$) | $k_m/k_s$ | Ref. |
|---|---|---|---|---|
| Polyimide | 27 | 0.1919 | 0.32 | [61] |
| Polyethylene terephthalate | 20 | 0.25 | 0.43 | [64] |
| Polycarbonate | 23 | 0.23 | 0.38 | [65] |
| Nafion 117 | 17~65 | 0.13~0.29 | 0.217~0.483 | [42, 66] |
| Silica | 0~100 | 1.4~1.6 | 2.33~2.67 | [61] |

**Acknowledgments**

Funding: This work was supported by the National Natural Science Foundation of China (grant numbers 51976157, 51721004), the funding for selected overseas talents in Shaanxi Province of China (grant number 2018011), the Fundamental Research Funds for the Central Universities (grant number xzy012020075) and the China Scholarship Council (grant number 202106280109).

**CRediT authorship contribution statement**

**W.Z.**: Conceptualization, Investigation, Methodology, Software, Formal analysis, Validation, Visualization, Writing - Original Draft, Writing - Review & Editing. **M.F.**: Validation, Writing - Review & Editing. **K.J.**: Validation, Writing - Review & Editing. **F.Q.**: Writing - Review & Editing. **P.G.**: Writing - Review & Editing. **Q.W.**: Writing - Review & Editing. **C.Y.**: Writing - Review & Editing, Supervision. **C.Z.**: Conceptualization, Writing - Review & Editing, Supervision, Funding acquisition.

**Declaration of Competing Interest**

The authors declare that they have no known competing financial interests or personal relationships that could have appeared to influence the work reported in this paper.

**Data Availability Statement**

The data and codes that support the findings of this study are available from the corresponding author upon reasonable request.